\documentclass{aa}  

\usepackage{subfigure}
\usepackage{subfloat} 
\usepackage{float}
\usepackage{graphicx}
\usepackage{natbib,twoopt}
\usepackage[varg]{txfonts}
\usepackage{booktabs}
\usepackage[english]{babel}
\usepackage{txfonts}
\usepackage{soul}
\usepackage{tablefootnote}
\usepackage[dvipsnames]{xcolor}
\usepackage{tabularx}
\usepackage{booktabs}
\usepackage{ulem}

\begin{document} 

   \title{Gravity modes and potential evidence for Rossby Waves in late O-type supergiants}
 
   \author{Lydia S. Cidale\inst{1,2}\fnmsep\thanks{Member of the Carrera del Investigador Científico,  CONICET, Argentina}, Alejandra Christen \inst{3}, Aldana Alberici Adam\inst{1,2}\fnmsep\thanks{Fellow of CONICET}, Suryani Guha\inst{4,5}, Alex Lobel\inst{6}, Gunther F. Avila Marín\inst{3},  
          Michaela Kraus\inst{4}, Alejandro H. Córsico\inst{1,2}$^{,\star}$, and Julieta Sánchez Arias\inst{4}
          }

   \institute{Instituto de Astrofísica de La Plata, Paseo del Bosque s/n, B1900FWA, La Plata, Provincia de Buenos Aires, Argentina\\
         \email{lydia@fcaglp.unlp.edu.ar}
             \and
  Facultad de Ciencias Astronómicas y Geofísicas, Paseo del Bosque s/n, B1900FWA, La Plata, Provincia de Buenos Aires, Argentina   
           \and
 Instituto de Estadística, Universidad de Valparaíso, Av. Gran Bretaña 1111, Casilla 5030, Valparaíso, Chile
              \and
Astronomical Institute, Czech Academy of Sciences, Fričova 298, 25165 Ondřejov, Czech Republic
              \and
Astronomical Institute, Faculty of Mathematics and Physics, Charles University, V Hole\v{s}ovick\'{a}ch 2, 182 00 Prague, Czech Republic
         \and
  Royal Observatory of Belgium, Ringlaan 3, 1180 Brussels, Belgium         
}

   \date{Received August, 2024; accepted  2024}
    
  \abstract 
   {The physical properties of O-type supergiant stars remain largely unexplored. By analysing their light variations, we can sometimes unravel underlying physical phenomena, such as binarity, stellar pulsations, and rotation modes, as well as discover previously unobserved or unexplored phenomena in these types of stars.}
   {This study aims to analyse the TESS light curves of three O-type supergiant stars to identify periodic signatures and gain deeper insights into their internal dynamics and structure. Our primary goal is to search for the presence of rotational modulation and possible evidence of Rossby waves.}
   {A period search was performed on photometric data of three supergiant O-type stars using the generalised Lomb-Scargle periodogram and the weighted wavelet Z-transform together, leveraging the advantages of each method. We explored phase diagrams and searched for rotational splitting. If the observed wave frequencies were consistent with the dispersion relation of global Rossby waves, we identified them as potential signatures of this mechanism. We used a graphical method to compare the observed and predicted frequencies.}
{The three analysed supergiant stars (HD~188001, HD~192639, and HD~159952) exhibit a comparable set of frequencies. We classify most of the observed frequencies as $g$-mode oscillations (either $\ell=1$ or $\ell=2$), in agreement with the evolutionary state of the objects. In all cases, we find evidence of rotational modulation. The remaining oscillation patterns may be consistent with Rossby ($r$) modes, including both tesseral and sectoral configurations. The angle of inclination of the rotation axis was estimated using stellar parameters available in the literature, together with the rotational period derived in this work. We also discuss a possible connection between the observed low-frequency waves and the red-noise component commonly observed in the periodogram of photometric time series of massive supergiants.
}
   {Our results provide evidence of $g$-mode oscillations modulated by rotation. In addition, we find that Rossby waves may be excited in rotating O-type supergiants, which suggests that these large-scale inertial oscillations could play a role in the observed low-frequency variability of such stars. A comprehensive theoretical treatment addressing the development of pulsations in rotating stars lies beyond the scope of the present work and is deferred to future studies.
   Further investigation would provide insight into the star’s surface flows, rotation, and angular momentum transport.} 
   

\titlerunning{Rossby Waves}
\authorrunning{Cidale et al.}
\maketitle

\newcommand{\sg}
\nolinenumbers 

\section{Introduction}

Global Rossby waves (or $r$ modes; hereafter RWs) are large-scale wave-like patterns or structures that naturally occur in a rotating fluid spheroid and extend over a substantial latitude range \citep[see][]{Rossby1939}. These low-frequency waves have recently been detected in many main-sequence stars using light curves collected from space missions \citep{VanReeth2016, Saio2018, Li2019,  Jeffery2020,  Samadi-Ghadim2020, Takata2020b, Saio2022}. 

Evidence of low-frequency waves, such as inertial gravity waves or internal gravity waves in general, was found in a limited number of blue supergiants, based on studies of their line-profile variability \citep{SimonDiaz2017, SimonDiaz2018, Aerts2018} or through their (stochastic) signals in high-precision, high-cadence light curves \citep{Bowman2019}, collected with the Kepler space telescope (K2 mission) and the Transiting Exoplanet Survey Satellite (TESS). However, to date, no systematic studies have been carried out to search for rotational modulation and the presence of RWs in massive supergiant stars.

Rossby waves are associated with the Coriolis force and arise due to the conservation of absolute vorticity on a rotating sphere. They cause periodic variations in surface pressure (temperature, density, or both), which may lead to the periodic modulation of the stellar radiance \citep{Saio2018}. 
Contrary to inertia-gravity waves (IGWs), the frequency of RWs is smaller than the angular frequency of rotation. The RW solution may have prograde motions and is characterised by the Rossby parameter, $\beta$, the zonal wave number, $n$, and the meridional wave number, $m$. The parameter $\beta = 2\Omega\, (\cos\,\theta)/R_\star$ is the latitudinal variation of the Coriolis acceleration, $\Omega$ and $R_\star$ are the angular rotation rate and radius of the star, respectively, and $\theta$ denotes latitude \citep{Zaqarashvili2021}.

\citet{Albekioni2023b} studied RWs in uniformly rotating stars with outer radiative zones. They find that these waves are confined to narrow surface layers and that their frequencies depend on stellar rotation, radius, and surface temperature. Studies of rapidly rotating stars revealed that the RWs are strongly trapped around the equator \citep{Townsend2003, Albekioni2023a}.

Photometric and spectroscopic observations of OB supergiants usually reveal complex quasi-periodic oscillations or irregular variations. Their light curves contain a mix of frequencies, reflecting different physical processes \citep[pulsation modes, rotational modulation, beat frequencies, stochastic low-frequency variability, and other phenomena; see for example][]{Burssens2020, Kourniotis2025}.
 Thus, the frequencies deduced from the light variations represent an essential source of information. They can help constrain stellar evolution models, in particular, because the structure of the pulsation modes and their frequency separations may yield physical parameters of the star, such as the rotation period or the composition of its layers.

This work analyses high-precision photometric observations from the TESS mission of three late O-type supergiant stars. Their spectra were modelled by \citet{Gormaz2022}, based on self-consistent wind solutions. These stars have high mass-loss rates that significantly impact their surroundings and exhibit substantial spectral and brightness variability. From an observational point of view, our ultimate goal is to detect signals in high-cadence photometric time series that could be interpreted as RWs, since these manifest themselves in the low-frequency domain and may be linked to semi-regular variability, due to wave interactions \citep{Zaqarashvili2021}. Furthermore, $r$-mode waves can offer a powerful diagnostic tool for probing the star's rotational profile \citep[cf.][]{Aerts2021}  and exploring the internal structure of luminous blue supergiants. 
Current models of massive stars often lack detailed knowledge about wave-induced processes, such as angular momentum transport and mixing in stellar interiors. Gaining this knowledge is essential for developing more accurate stellar evolution models.

This work is organised as follows. Section~\ref{Observations} presents the target selection and light curve extraction procedure.
Section~\ref{sec:methods}  outlines the techniques for searching for and identifying periods in uneven data samples. Section~\ref{pulsation} briefly summarises the typical stellar pulsation properties. Section~\ref{Results} presents the results obtained in the period search from time-series light curves and their analysis for each star under study. 
Sections~\ref{Discussion} and \ref{Conclusions} provide a discussion and our main conclusions. The Appendix~\ref{Ap:A} presents the light curves and frequency analysis.
 
\section{Target selection and observations}
\label{Observations}

We focused our study on three specific targets:  \object{HD~188001}, \object{HD~192639}, and \object{HD~195592} taken from \citet{Gormaz2022}.
We selected these O-type supergiants (with luminosity class Ia/Iab) because they are (rapidly) rotating and hence ideal candidates for exhibiting evidence of RWs, potentially detectable through the variability of their photometric light curves.

Table~\ref{tab_1} lists the stellar parameters of the three selected stars, such as the effective temperature ($T_{\rm eff}$), surface gravity ($\log\,g$), stellar radius ($R_\star$) and projected rotation velocity ($v\, \sin\,i$). We adopted the values reported in \citet{Gormaz2022} and \citet{Holgado2022}. This star sample is quite homogeneous with stellar radii distributed around $21$~R$_\sun$. Since the inclination angle of the rotation axis is unknown, we estimated an upper limit to the rotation period for an equator-on configuration ($i= 90^\circ$). We adopted typical errors for $\Delta T_{\rm eff} =1\,000$~K, $\Delta R_\star= 1$~R$_\sun$, and $\Delta\, v \sin \,i=10$ km~s$^{-1}$. This table helped us search for frequencies associated with rotational modulations.
 
\begin{table}[h]
\centering
\caption{Stellar parameters and the expected rotation period for an equator-on configuration.}
\label{sparam}
\resizebox{\columnwidth}{!}{
\begin{tabular}
{@{}rccccl}
\hline
$HD$~~~& $T_{\rm eff}$ & $\log\,g$ & $R_\star$ & $v\sin~i$ & ~${\it P_{\rm rot}}$~\rm[days]~ \\ 
& [kK] & & [R$_{\odot}$] & [km~s$^{-1}$] & $~~~i=90^\circ$ \\ 
\hline
\rule{0pt}{1em}%
$188001$ & $34.5$ & $3.32$ & $23.0$ & $90 $ & $12.9\pm1.5$  \\
$192639$ & $34.0$ & $3.25$ & $19.8$ & $82^{a}$  & $12.2\pm1.6$ \\
$195592$ & $29.5$ & $3.20$ & $21.5$ & $60$ & $18.1\pm3.1$ \\
\hline
\end{tabular}
}

\label{tab_1}
\begin{flushleft}
\tiny{Data taken from \citet{Gormaz2022}  and
$^{a}$\citet{Holgado2022}.}
\end{flushleft}
\end{table}

The TESS mission \citep{Ricker2014}, launched by NASA, observes the sky in sectors measuring $24^{\circ}\times 96^{\circ}$.  Each sector is observed for two satellite orbits around the Earth, of about $27$~days on average.
TESS provides high-cadence photometric time series for in-depth analyses of stellar variability.

 The light curves of all our targets were observed from July 2018 to January 2024. Observations per object range from three to five sectors, with a maximum of two consecutive sectors. The light curves were retrieved from the Mikulski Archive for Space Telescopes (MAST) database\footnote{https://mast.stsci.edu/portal/Mashup/Clients/Mast/Portal.html}.
 For each target, we extracted  $20\times20$ pixel cutouts from the TESS full-frame images (FFIs) across all available sectors. These cutouts contain time-series data sampled at 30-minute intervals. Since the Science Processing Operations Center (SPOC) does not provide predefined photometric apertures for FFI cutouts, we used custom aperture photometry techniques to extract light curves from the images. The \texttt{'Lightkurve'} \citep{Lightkurve} Python module was used to process the data and extract the light curves. To mitigate contamination from nearby background and foreground stars, we utilised Gaia Data Release~$2$ (DR2) \citep{Gaia2} to accurately determine the positions and magnitudes of stars within the field of view. Apertures for the targets were manually selected with precision, avoiding any contamination. To measure background flux, background pixels were selected after target masking. To extract the light curves, the background flux was deducted from the flux of the target mask. The fluxes and their errors were converted into magnitudes and then normalised to the mean value of the entire light curve for each sector. The time-series data were analysed following the procedure described in Sect.~\ref{sec:methods}.
   
\section{Time series analysis methods for period detection}
   \label{sec:methods}
Time-frequency methods (e.g. wavelet analysis), which obtain the frequencies detected at each time instant, are effective when using methods based only on average frequency detection, such as periodograms, which, by eye, lead to failure. However, non-uniform sampling is a general problem in astronomical observations, for example, when dealing with light curves of variable stars.
To address this issue, we selected two methods, the generalised Lomb-Scargle 
(GLS) periodogram and the weighted wavelet Z-transform (WWZ) to assess as many periods as possible. Later, we averaged the periods obtained to determine a more accurate value because it combines the strengths of each method. 

\citet{Han2012} analysed the pros and cons of all these methods. These authors argue that the wavelet transforms are capable of multi-resolution analysis and can be analysed in local time and frequency domains for evenly sampled time series. They add that results on the periodicities contained in the light curve can be obtained from the average power plot (similar to the periodogram displaying the frequencies and their wavelet power) and from the scalogram, which shows the frequency change during the observed time interval. However, when applied to an unevenly spaced one, the response of the discrete wavelet transform (DWT) strongly depends on the local number density of the observational data points. Consequently, the analysis may show a time-false evolution of periods and detect spurious high frequencies.
Moreover, \citet{Alberici2023} discussed the search for periods provided by the Lomb-Scargle (LS) and classic wavelet analysis using simulated light curves.  These authors recommend applying various methods to analyse TESS light curves to search for periods. They show that Morlet wavelet analysis produces a systematic shift in the resulting periods, underestimating their values. Therefore, they suggest using the WWZ \citep{Foster1996}, an improvement of wavelet analysis to cover an unevenly sampled dataset.
Due to similar problems in period detection, Fourier analysis was adapted to irregularly sampled data. The most commonly used method with these characteristics -- based on Fourier analysis -- is the LS periodogram. However, it has certain limitations, for example, in the presence of non-sinusoidal signals, large data gaps, red noise, leading to spurious peaks \citep{VanderPlas2018}.
 The following subsections summarise the methods used for performing the frequency analysis. 

\subsection{The Lomb-Scargle periodogram}

The LS periodogram comes from the discrete Fourier transform \citep[DFT;][]{Lomb1976, Scargle1982}, in which a time series is decomposed into a linear combination of sinusoidal functions characterised by the frequency $f$, where the angular frequency is $w=2\pi\,f$.
Scargle modified the standard periodogram formula to find a time delay, $\tau$,  which is defined by
\begin{equation*}
\tan 2w\,\tau=\frac{\sum_{j} \sin 2w\,t_j}{\sum_{j} \cos 2w\,t_j},
\end{equation*}

\noindent such that the pair of sinusoids would be mutually orthogonal at sample times $t_j$ (the observed time). 
Moreover, fine-tuning the values used to enhance the estimate of power at a frequency, $w$, ensures that the transform remains invariant to time shifts when employing two basis functions. In this context, it shares the same statistical distribution as the periodogram in evenly sampled scenarios. Based on N pairs of observations $(t_i, x_i)_{1\leq i\leq N}$, the periodogram \citep{VanderPlas2018} is computed by

\begin{equation}\nonumber
	P_{LS}(w)=\frac{1}{2}\left(\frac{\left(\sum_{j}x_j \cos w\,(t_j-\tau)\right)^2}{\sum_{j} \cos^2 w\,(t_j-\tau)}+\frac{\left(\sum_{j}x_j \sin w\,(t_j-\tau)\right)^2}{\sum_{j} \sin^2 w\,(t_j-\tau)} \right).
\end{equation}

\noindent \citet{Zechmeister2009} improved the formula by considering a weight that takes into account the measurement errors of the data, and obtained the GLS periodogram. According to the authors, the generalised approach may offer advantages, as it can produce more accurate frequency estimates, reduce susceptibility to aliasing, and allow for a more reliable assessment of spectral intensity.



\noindent Here, we utilised the \textsc{timeseries.lombscargle} package from the \textsc{python} module \textsc{astropy} to compute the GLS periodogram. The resulting power spectral density (PSD) is a dimensionless value within the interval $[0, 1]$, following the standard normalisation outlined by \citet{Zechmeister2009}.
The frequency analysis was performed over the range from $f_{\mathrm{min}}$ to $f_{\mathrm{Nyq}}$, where $f_{\mathrm{min}} = 1 / \Delta T$, $f_{\mathrm{Nyq}}$ is the Nyquist frequency \citep{Lenz2005}, and $\Delta T$ is the total time span of the data. No significant frequency above $1.5~\mathrm{d}^{-1}$ was detected in any of our objects. 

An iterative pre-whitening process was applied to achieve greater precision in identifying significant frequencies. The frequency with the highest amplitude was identified in each iteration, and the corresponding sinusoidal model was fitted and subsequently removed. The residuals obtained from this fit were then used in the next iteration to extract the following frequency. This process was repeated until the signal-to-noise (S/N) ratio of three consecutive extracted signals fell below a selected S/N limit. According to \citet{Baran2021}, to reduce the likelihood of detecting spurious periods in TESS data, it is recommended to adopt a threshold of $\mathrm{S/N} \ge 5$, where $S/N$ is defined as the ratio between the amplitude of the extracted frequency and the average amplitude of the residual spectrum within a symmetric frequency window centred on the extracted frequency \citep{Bowman2019, Bowman2020}.  Since RWs may occur within the same range as stochastic low-frequency variability, this criterion could potentially exclude several relevant frequencies in those regimes.
Therefore, we adopted the classical and more flexible cut-off of $\mathrm{S/N} \ge 4$, where the noise is calculated using a frequency window of $5~\mathrm{d}^{-1}$ (see e.g. \citealt{Breger1993, Burssens2020}).

\subsection{The weighted wavelet Z-transform}

The WWZ transform was proposed by \citet{Foster1996} to detect periodicities specifically in unevenly spaced light curves. In this context, the wavelet is re-scaled to satisfy the admissibility condition on such irregular sampling. To achieve this goal, the signal is projected onto three trial functions -- $\varphi_1(t)=1$,
$\varphi_2(t)=\cos(w\,(t-\tau))$, and
    $\varphi_3(t)=\sin(w\,(t-\tau))$ -- related to the wavelet Morlet given by the base function $f(t)=e^{i w (t-\tau )-c\,w^2 (t-\tau )^2}$, to compensate for the missing data due to uneven time spacing\footnote{Here, $w=2 \pi f$, $\tau$ is the time shift, and $c$ is the decay rate.}. In addition, a model function $y(t)$, defined by $y(t)=\sum_{\alpha} y_\alpha \varphi_\alpha (t)$, is calculated from the data using least squares, and the values of both approaches are compared through a variance estimation which is linked with the wavelet power computation.

For each frequency, the WWZ method compares the weighted variation of the data, $V_x$, with the weighted variation of the model, $V_y$, identifying as significant those in which these two quantities are similar. The weighted variations with respect to the signal, $x(t)$, are then given by 
\begin{equation*}
V_x=\frac{\sum_{\alpha} w_\alpha x^2(t_\alpha)}{\sum_{\lambda}w_\lambda}-\Big(\frac{\sum_{\alpha} w_\alpha x(t_\alpha)}{\sum_{\lambda}w_\lambda}\Big)^2,
\end{equation*}
and
\begin{equation*}
V_y=\frac{\sum_{\alpha} w_\alpha y^2(t_\alpha)}{\sum_{\lambda}w_\lambda}-\Big(\frac{\sum_{\alpha} w_\alpha y(t_\alpha)}{\sum_{\lambda}w_\lambda}\Big)^2,
\end{equation*}
where $w_{\alpha}=e^{-cw^2(t_\alpha-\tau)^2}$ for $\alpha=1, \cdots, N$ are the weights corresponding to the Morlet wavelet bases and $N$ is the size of the data.

Formally speaking, the WWZ for the signal $x(t)$ is given by

\begin{equation} 
WWZ(x)=\frac{(N_{eff}-3) V_y}{2(V_x-V_y)},
\label{EQ:WWZ}
\end{equation}

\noindent where the effective number $N_{eff}$ is defined by
$N_{eff}:=~\left(\sum_{\alpha} w_\alpha\right)^2/\sum_{\alpha}w_\alpha^2$.
To determine the power of the statistic WWZ, we used the \textsc{libwwz} package\footnote{https://github.com/ISLA-UH/libwwz.git} in \textsc{python}. The method is based on picking the peak of WWZ to derive the true period \citep[see details in][]{Foster1996}. All frequencies displaying a maximum peak in the average wavelet power spectrum (time-averaged frequency) were selected for further consideration (see e.g. Fig.~\ref{fig:lightcurve9sge}), and the scalogram (frequency vs time, where the scale is inversely related to frequency) was evaluated to derive the interpretation of the results. 

\subsection{Frequency uncertainties}\label{uncert}

For both techniques, the uncertainties were estimated through $1000$ Monte Carlo simulations, which incorporated the intrinsic magnitude errors of the TESS measurements. The frequency error is usually lower than $10^{-5}$~d$^{-1}$. To identify independent frequencies, we adopted the criterion proposed by \citet{Loumos1978}, in which the differences between two observed  values exceed the more restricted resolution criterion of $1.5/\Delta~T$, where $\Delta T$ is the total time span of the TESS sectors selected. This value changes according to the number of contiguous sectors in time that are selected.

\section{Pulsation modes}
\label{pulsation}

The global oscillations of a star have a discrete spectrum of frequencies characterised by radial and non-radial pulsation modes, which provide insights into a star's internal structure. Radial modes involve symmetric pulsations, while non-radial modes feature complex distortions. \citet{Saio2018} reviewed the principles of adiabatic pulsations and excitation mechanisms, including the $\kappa$ mechanism, turbulent convection, and the $\epsilon$ mechanism, highlighting their dependence on stellar properties and evolutionary stages. Non-radial spheroidal pulsations consist primarily of $p$ modes (pressure modes) and $g$ modes (gravity modes). $p$ modes typically have higher frequencies than $g$ modes. These frequencies systematically increase with radial order ($k$), yielding a nearly regular spacing between consecutive overtones, particularly for high-order $p$ modes. In contrast, the frequencies of the $g$ modes decrease as the number of nodes increases, and the separation of periods of high-order $g$ modes is equidistant. 

\subsection{The $r$ modes}
The $r$ modes (toroidal modes) are prominent in stars with significant rotational velocities. They include low-frequency rotational waves, (i.e. RWs) -- which are influenced by the Coriolis force and associated with pressure changes -- and high-frequency gravity waves, (i.e. IGWs), where buoyancy serves as the restoring force.
\citet{Haurwitz1940} \citep[cf.][]{Zaqarashvili2021} showed that the conservation of total vorticity over a 2D spherical surface with a non-divergent character leads to the solution for the stream function in terms of associated Legendre polynomials with the dispersion relation,

\begin{equation}
    \sigma = -\frac{2m\,\Omega}{\ell\,(\ell+1)},
    \label{Rossby_formula}
\end{equation}

\noindent where $\Omega$ is the rotational rate of the system. Moreover, $m$ and $\ell$ are integers ($\|m\|\leq \ell$) and represent the angular order and degree of associated Legendre polynomials, respectively \citep[see][]{Zaqarashvili2021}.  The difference $\ell~-~\|m\|$ determines the number of zeroes (nodes) between the north and south poles. Spherical harmonics with $\ell=\|m\|$ are zonal harmonics, and those with  $\ell\neq \|m\|$
 are tesseral harmonics. 

However, within the zeroth-order approximation, this relation does not determine the radial structure of the modes.  \citet{Papaloizou1978} estimated the correction to the previous RW dispersion relation, due to the aspherical stellar form for high degree modes ($\ell >1$). In the approximation of slow rotation \citep{Provost1981}, the dispersion relation is 
\begin{equation}
 \sigma=\sigma_0\,\left[1+\left(\frac{\Omega}{\Omega_g}\right)^2 \sigma_1\right], 
 \label{Eq_correction}
 \end{equation}

\noindent where $\Omega_g=\sqrt{GM_\star/R_\star^3}$ is the characteristic frequency of the star, $\sigma_0$ is the zeroth-order eigenfrequency, and $\sigma_1$ 
denotes the first-order rotational correction. The corresponding eigenvalues ($\sigma_1$) for a polytropic stellar model are given by \citet{Provost1981}. In the present study, the correction to the RW dispersion relation has only a minor impact on the $r$-mode frequencies, since $(\Omega/\Omega_g)^2$ is of the order of $1.5\%-2.0\%$ (as derived from Table~\ref{sparam}). In consequence, we adopted the zeroth-order approximation throughout this work.

 Then, in the inertial frame, assuming uniform rotation, the dispersion relation for $r$ modes is given by

\begin{equation}
    \sigma = -\frac{2m\,\Omega}{\ell\,(\ell+1)} + m\,\Omega.
    \label{Inertial frame}
\end{equation}
  
\noindent When $\ell$ increases, the r-mode frequencies are of the order of $\sigma\approx m~\Omega$. However, assuming differential rotation, \citet{Papaloizou1978} showed that Eq.~(\ref{Inertial frame}) also admits the possibility of Kelvin-Helmholtz instability, leading to a more complex situation. 

\subsection{The stellar evolution code}

{\tt LPCODE} is a 1D stellar evolution code that was initially developed to model the complete evolutionary sequence of low- and intermediate-mass stars, from the zero-age main sequence through advanced evolutionary stages \citep{Althaus2005, 2016A&A...588A..25M}. 
It solves the standard stellar structure equations (mass conservation, hydrostatic equilibrium, energy transport, and energy conservation) using a Henyey-type relaxation scheme. {\tt LPCODE} uses radiative opacity taken from the OPAL database. The adopted nuclear network includes the main burning processes (pp chains, CNO cycle, helium burning, and advanced reactions), allowing for a consistent treatment of energy generation and chemical evolution. Time-dependent mixing processes are modelled through a diffusion approach, including convection (treated within the mixing-length theory), convective overshooting, and diffusion (gravitational settling, chemical diffusion, and thermal diffusion).

\begin{figure}[h!]
    \centering
\includegraphics[width=0.9\linewidth]{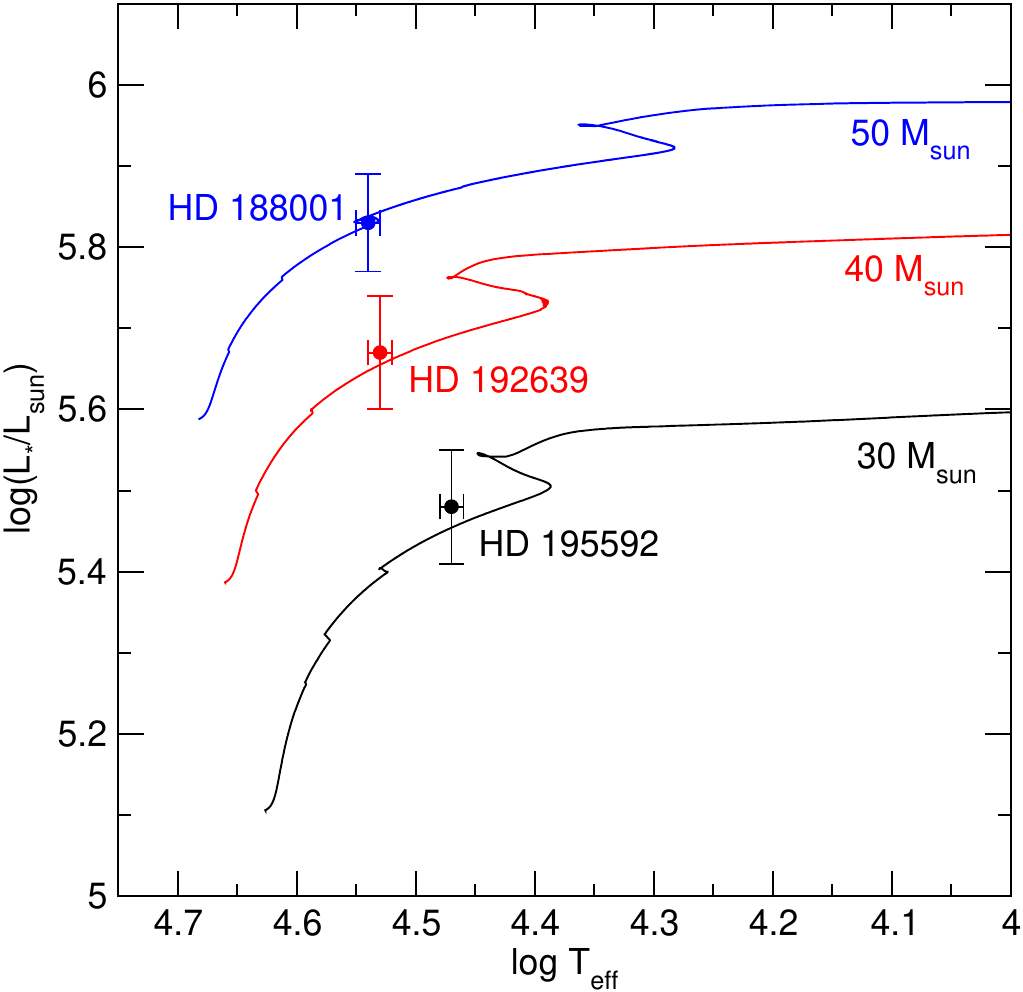}
    \caption{Hertzsprung-Russell diagram showing {\tt LPCODE} evolutionary tracks for stars with stellar masses $M_{\star}= 30, 40$, and 
    $50$ M$_{\sun}$. Symbols with error bars correspond to the location of the stars \object{HD~188001}, 
    \object{HD~192639}, and \object{HD~195592}.}
\label{fig:HR-LPCODE}
\end{figure}

To identify the pulsation modes, we computed an equilibrium stellar model for each star by adopting the stellar luminosity inferred from the parameters listed in Table~\ref{tab_1} and selecting the corresponding evolutionary track (see Fig.~\ref{fig:HR-LPCODE}). We adopted a metallicity of $Z= 0.01$, a mixing length parameter using the mixing length theory (MLT) of convection $\alpha_{\rm MLT}= 1.822$, and convective overshooting with $f_{\rm OV}= 0.0174$, where $f$ is a measure of the extent of the overshoot region. We computed dipole ($\ell= 1$) and quadrupole ($\ell= 2$) non-radial $p$ and $g$ pulsation modes with the help of the pulsation code {\tt LP-PUL} \citep{2006A&A...454..863C}, which solves the equations governing the linear, adiabatic, non-radial stellar oscillations. This pulsation code has been extensively employed in asteroseismological analyses of pulsating white dwarf stars \citep{2019A&ARv..27....7C}.

For our sample of O-type stars, the p modes have periods shorter than one day. Therefore, Table~\ref{theoretical_comparison} lists only the non-radial $g$-mode oscillations ($\ell =1$ and $\ell=2$) used for comparison with the observations. Since no uniform period spacing is observed between consecutive modes, these modes are not in the asymptotic regime.\footnote{The asymptotic regime of non-radial pulsation modes corresponds to the limit in which the radial order of the mode is large ($k \gg 1$). In this regime, the separation of the periods of $g$ modes with consecutive radial order is constant and is described by approximate analytical expressions  \citep{1980ApJS...43..469T}.}
 
\section{Results}
\label{Results}
We present light curves of three O-type supergiants, based on observations of three (\object{HD~188001}), five (\object{HD~192639}), and five (\object{HD~195592}) TESS sectors, and analysed with the LS and WWZ techniques. The total number of significant detected frequencies corresponding to each star is presented in the appendix (Tables~\ref{table:188001}, \ref{table:HD192639}, and  \ref{table:HD195592}).
 Identified frequencies from different sectors were grouped and subsequently averaged when their mutual separations were smaller than the minimum frequency difference defined by the Rayleigh resolution criterion  ($1/\Delta T$; see Sect.~\ref{uncert}). The last two columns of the tables list the mean frequencies obtained from both methods, followed by their dispersions and the associated mean period.
To search for independent frequencies and assess possible restoring pulsation mechanisms, we verified that the resolving power exceeded the more restrictive Rayleigh criterion \citep[$1.5/\Delta T$,][]{Ramirez2025}, ensuring that closely spaced frequencies could be reliably distinguished. 

 We identified the $g$-mode oscillations using the results summarised in Table~\ref{theoretical_comparison}. Possible $r$-mode oscillations were investigated by applying the dispersion relation of Eq.~(\ref{Inertial frame}).
 The following subsections present the time-frequency analysis of the individual objects. 
 
\subsection{Analysis of HD~188001}

\object{HD~188001} (\object{9~Sge})  is a runaway star \citep{Underhill1995} of spectral type O7.5~Iab \citep{Sota2011} with $T_{\rm eff}~=~34\,000$~K, $\log\,g = 3.36$, and $R_\star = 20.8$~R$_\sun$, according to the catalogue of \citet{Martins2005}. Recently, \citet{Gormaz2022} estimated their stellar parameters, reporting a $T_{\rm eff} = 34\,500$~K, $\log\,g =~3.32$, $R_\star= 23$~R$_\sun$, and~$v\sin~i~=~90$~km~s$^{-1}$. From these values, the expected rotational period of \object{HD~188001} must be shorter than $13$~days (see Table~\ref{sparam}). \citet{Maiz2019} report periodic radial velocity variations of $97.6$~days. 

\begin{figure}[h!]
    \centering
\includegraphics[width=0.9\linewidth]{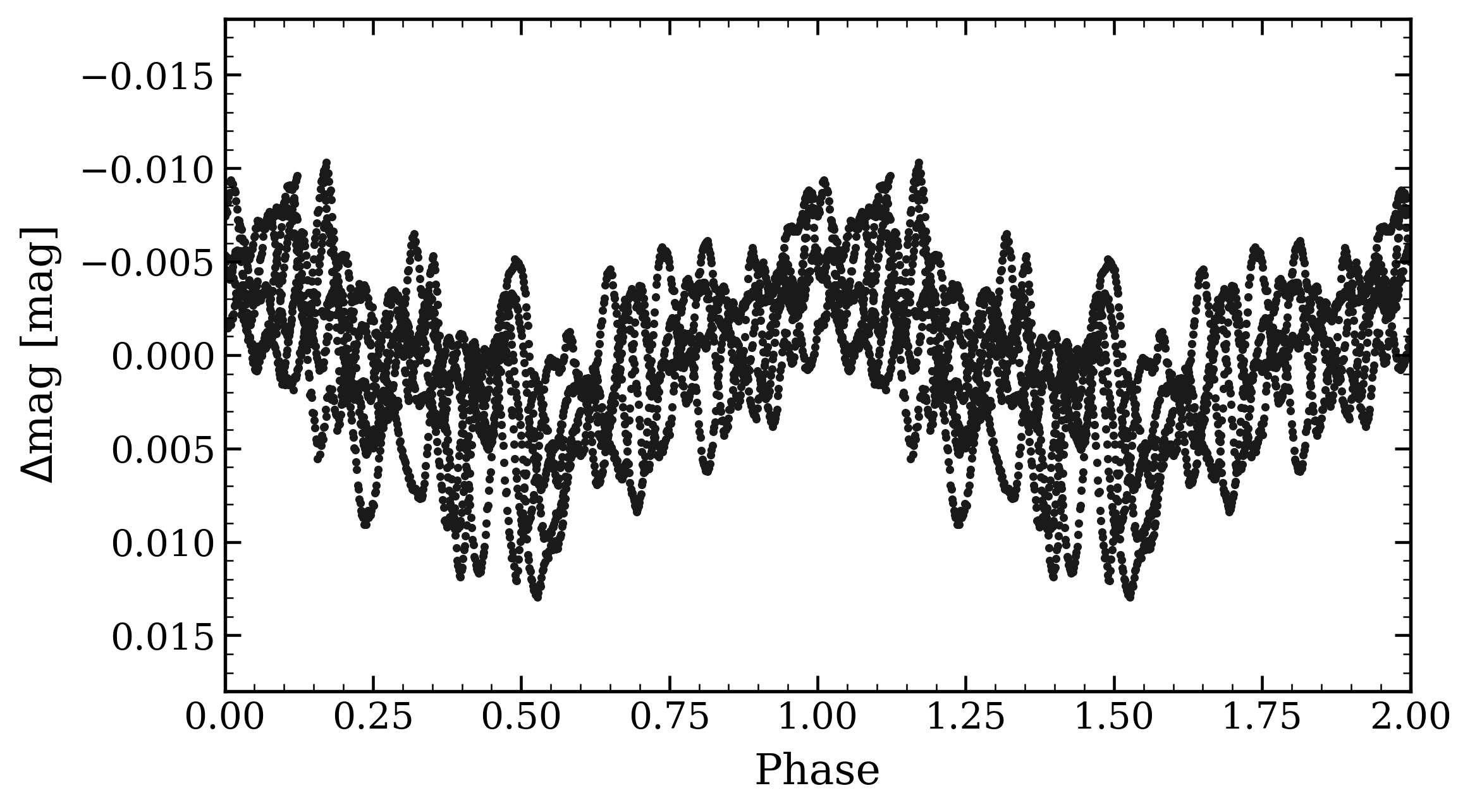}
    \caption{Phase-folded diagram of HD 188001 obtained from light curves using TESS sector $41$. A coherent physical signal is recovered using a period of $P=4.963$~d.}
\label{fig:PD_HD188001}
\end{figure} 

\begin{figure}[h!]
    \centering
\includegraphics[width=0.8\linewidth]{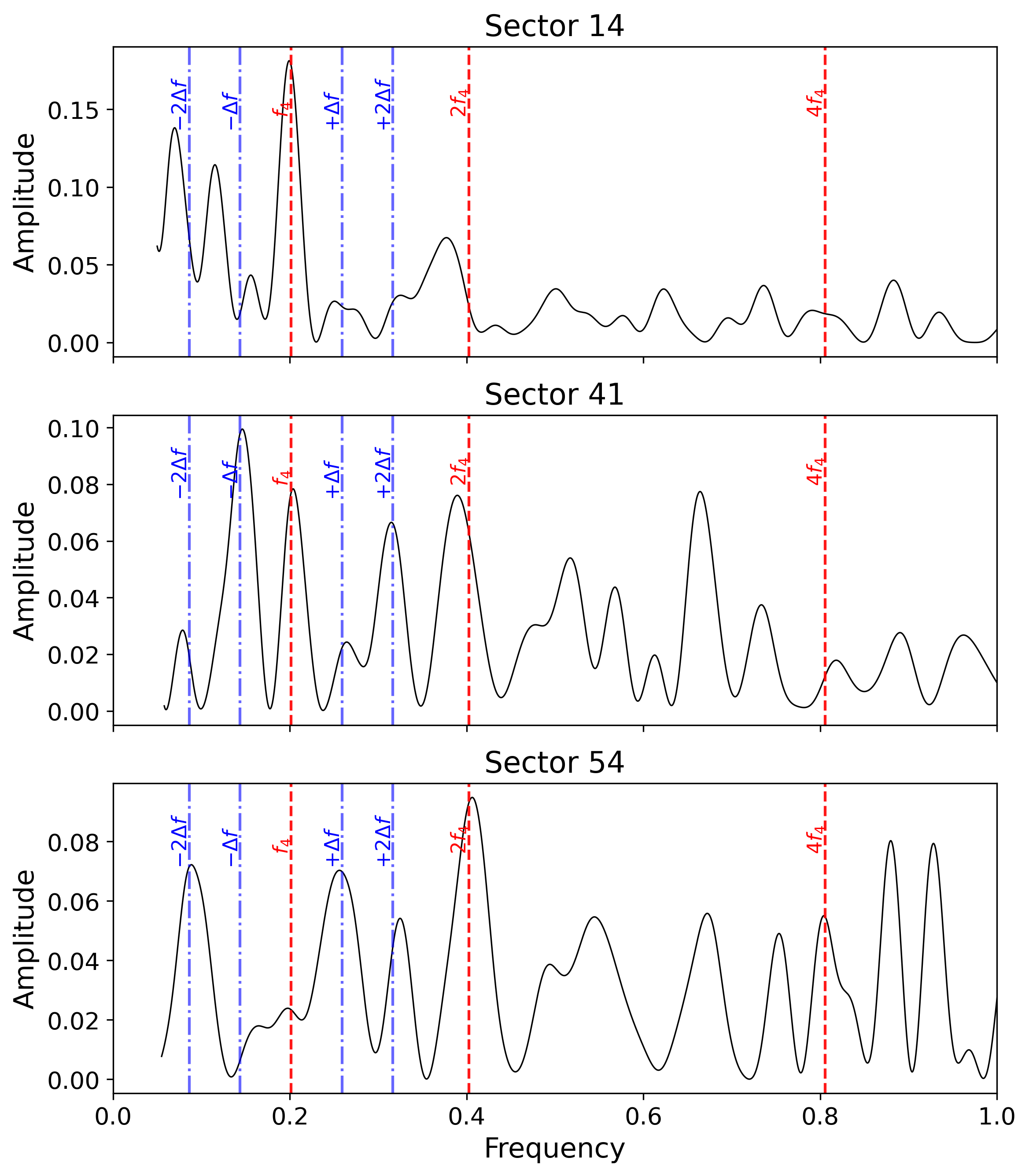}
\caption{Periodogram for HD 188001 using $\Delta f= 0.0576$ (blue lines) centred at the frequency $f_4=0.2014$~d$^{-1}$. Harmonics of $f_4$ are indicated with red lines.}
    \label{fig:fr_hd188001}
\end{figure}

\begin{table}[h]
\caption{\object{HD 188001}: Analysis of selected average frequencies given in Table~\ref{table:188001}.}
\label{HD188001_analysis}
\tabcolsep 2.0pt
\begin{tabular}{clrcc|cc}
\hline
\hline
~$ID$ & \hspace{1cm}$f_{i}$ & ~$f_{i}-f_{i-1}$ & $P$  & ~$\Delta P$ & \multicolumn{2}{c}{$Comments$}~~ \\
 &\hspace{0.8cm}[d$^{-1}$]&[d$^{-1}$]~~~&[d] & [d]&&\\
\hline
1&$0.0816\pm0.0063$&& $12.25$&$3.46$&$f_4- 2\Delta f$&\\
2&$0.1137\pm0.0019$&$0.0321$&$8.79$&$1.89$& &$(3,2)$\\
3&$0.1449\pm0.0001$&$0.0312$&$6.90$&$1.94$&$f_4-\Delta f$ & $g-mode$ \\
4&$0.2014\pm0.0028$&$0.0565$&$4.96$&$1.11$& $f_4$& $g-mode$\\
5&$0.2597\pm0.0058$&$0.0583$&$3.85$&$0.67$&$f_4+\Delta f$ & $g-mode$\\
6&$0.3148\pm0.0046$&$0.0551$&$3.18$&$0.63$&$f_4+ 2 \Delta f$& $g-mode$\\
7&$0.3922\pm0.0094$&$0.0774$&$2.55$&$0.63$&$2\,f_4$& $g-mode$\\
8&$0.5206\pm0.0110$&$0.1284$&$1.92$&$0.41$&$2\,f_5$& $g-mode$\\
9&$0.6629\pm0.0041$&$0.1423$&$1.51$&$0.17$ &$2 f_5+f_3$ &$g-mode$\\
10&$0.7451\pm0.0046$&$0.0822$&$1.34$&$0.18$ & $3 f_4+f_3$&$g-mode$ \\
11&$0.8614\pm0.0073$&$0.1163$&$1.16$&$0.09$&$f_4+f_9$ &\\
12&$0.9313\pm0.0075$&$0.0699$&$1.07$&$0.14$&$3\,f_6$ &\\
13 & $1.0793\pm0.0003$ & $0.1480$ & $0.93$ & & $2\,f_8$ &\\
\hline
\end{tabular}
\tablefoot{We identify an approximately uniform frequency spacing of $\Delta f = 0.0576$~d$^{-1}$. The frequency $f_4=0.2014$~d$^{-1}$ is taken as reference. The identified $g$ modes and candidate $r$ modes, labelled as $(\ell,m)$, are listed in the last column.}
\end{table}

The TESS light curves of \object{HD~188001}, observed in sectors $14$, $41$, and $54$, are shown in Appendix \ref{Ap:A}, along with their WWZ scalogram, WWZ average power, and LS periodogram.
There are slight variations in the frequency values and power across the three sectors, as illustrated in the scalogram of Fig.~\ref{fig:lightcurve9sge}. 
The frequencies obtained from the pre-whitening analysis, along with their corresponding averaged values, are listed in Table~\ref{table:188001}. Among the dominant frequencies, we identify a weak periodic signal with a period of $4.963$~days, corresponding to a frequency of $0.2015$~d$^{-1}$, in excellent agreement with the mean frequency value of  $f_4=0.2014$~d$^{-1}$ listed in Table~\ref{HD188001_analysis}.  It leads to a coherent physical signal with an amplitude of approximately $10$ mmag, as illustrated in the phase-folded diagram in Fig.~\ref{fig:PD_HD188001}.
Since this signal is detected across all sectors, we interpret it as arising from a stable oscillation.

Interestingly, the frequency analysis of the TESS sectors (see Fig.~\ref{fig:fr_hd188001}) reveals a pattern characterised by an approximately constant frequency spacing of $\Delta f = 0.0576$~d$^ {-1}$, as derived from a mean-square optimisation. The frequency, $f_4$, was used as a reference to illustrate the pattern. This behaviour is shown in Table~\ref{HD188001_analysis}, which lists the dominant frequencies and provides the difference between the observed frequency and period peaks. No pattern of evenly spaced periods is detected. 

For this star, we infer a stellar mass of approximately $50$~M$_\sun$, as shown in Fig. \ref{fig:HR-LPCODE}. Then, from the theoretical adiabatic oscillation modes (given in Table~\ref{theoretical_comparison}), we find that $f_4$ is compatible with the theoretical predictions for both $\ell=1$ and $\ell=2$ $g$-mode pulsations. However, given the quintuplet pattern in the observed frequency spectrum, $f_4$ is more likely associated with an $\ell=2$ gravity mode.
Table~\ref{HD188001_analysis} also lists all $g$ modes identified through comparison between the observed and theoretical values, considering only frequencies that fall within the Rayleigh frequency resolution.
We remind readers that the absence of constant period spacing in both the model (see Table~\ref{theoretical_comparison}) and the observations indicates that the $g$ modes lie outside the asymptotic regime.

Although we note that this frequency separation is comparable to a more restrictive Rayleigh resolution criterion, 
$1.5/\Delta T~=~0.055$, it may tentatively be interpreted as rotational splitting. We note, however, that the frequencies, $f_1$ and $f_3$, could potentially be affected by instrumental effects, as the TESS orbital frequency is $0.074$~d$^{-1}$. 

Assuming that the observed frequency separation, $\Delta f$, arises from rotational splitting, we estimated the rotational frequency using the relation,
\begin{equation} \label{eq:gmode}
    \Delta f = m~(1-C_{l})\,f_{\rm rot},
\end{equation}
\noindent
where $m=1$, $C_{l}=0.166$  and $f_{\rm rot} $ the rotational frequency \citep{Dziembowski1992}. We obtained $f_{\rm rot}=0.0691$~d$^{-1}$, which corresponds to a rotation period of $14.5\pm4.2$~d, noting the limitation imposed by the low-frequency resolution. 

\begin{figure}[h!]
    \centering
\includegraphics[width=0.8\linewidth]{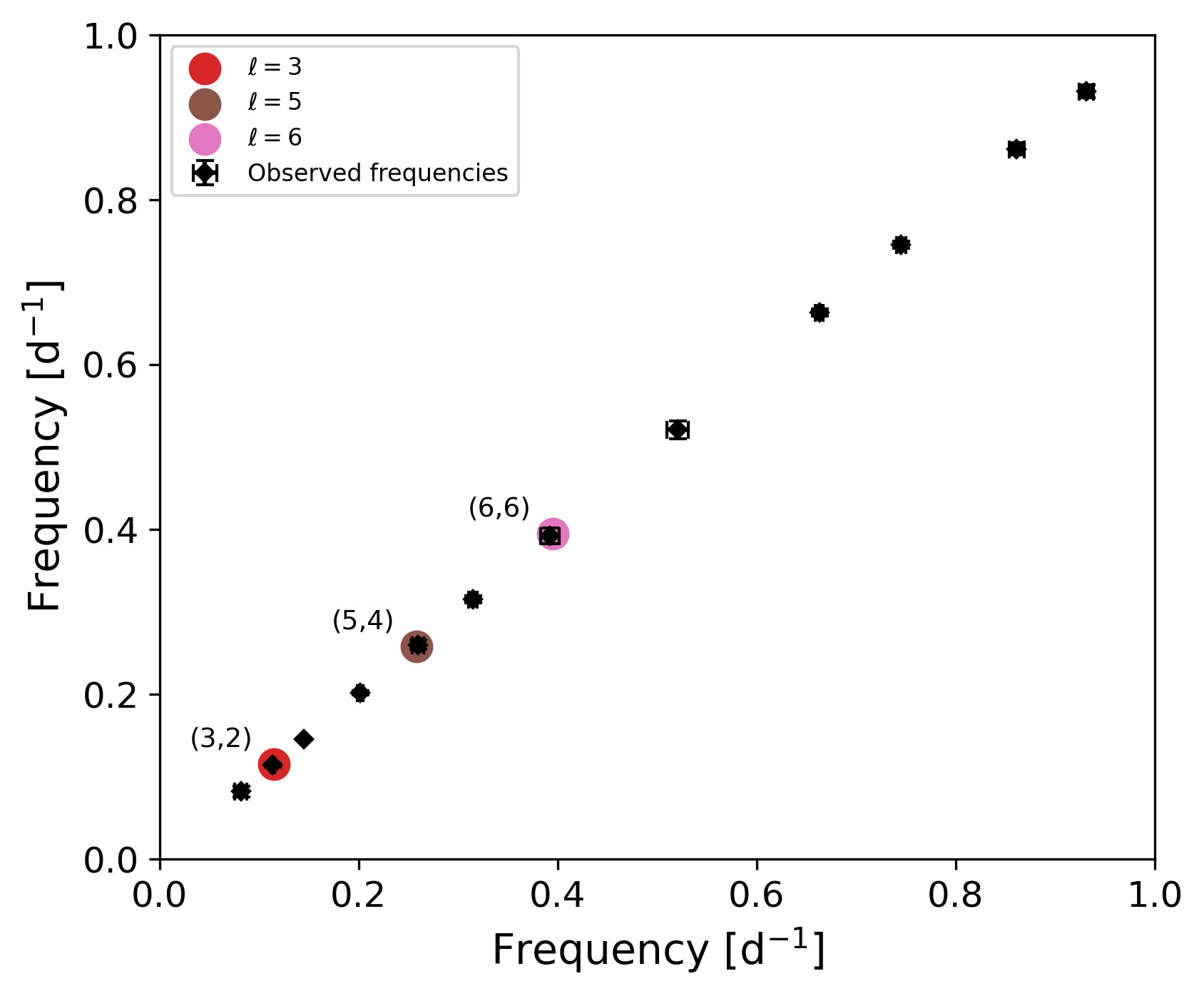} 
\caption{\object{HD~188001}: $r$-mode candidates obtained with the rotation frequency of $f_{\rm rot}=0.0691$~d$^{-1}$, using several values of $(\ell,m)$. Average observed frequencies, along with the standard deviation, are indicated by a black symbol.}
    \label{fig:Rossby_HD188001}
\end{figure}

The search for potential RWs was done using  $\Omega~=~2~\pi ~f_{\rm rot}$ in Eq.~(\ref{Inertial frame}) for various values of $\ell$ and $\|m\| \leq \ell$. The comparison between the observed and theoretical frequencies ($\sigma~=~2~\pi~f$), as described by the Rossby dispersion relation, is graphically presented in Fig.~\ref{fig:Rossby_HD188001} for different zonal $\ell$ values. As shown in Table~\ref{HD188001_analysis}, after removing the quintuples, harmonics, and frequency combinations, in addition to the $g$ modes, we found the independent frequency $f_{2}$. Since it is marginally resolved in sector $14$, it may be identified as a Rossby mode with $(\ell,m)=(3,2)$ (see Fig.~\ref{fig:Rossby_HD188001}). Such a mode is expected to become detectable at a rotation-axis inclination of approximately $60^\circ$, where the visibility of tesseral $r$ modes is maximised \citep{Saio2018}. The observed frequencies, $f_{5}$ and $f_{7}$, might blend with $r$ modes with $(\ell,m)=(5,4)$ and $(\ell,m)=(6,6)$. Although the visibility of these $r$ modes is very low, it might produce the beating-like effects observed in the phase-folded diagrams and lead to sector-to-sector variations in the observed power of these frequencies. Finally, the $97.6$-day period reported by \citet{Maiz2019} cannot be confirmed in this study due to the limited observation time of the light curves. Such a long period could suggest the presence of a companion.

 \subsection{Analysis of HD~192639}

 \object{HD~192639} is a blue supergiant star with a spectral type of O7.5~Iabf \citep{Sota2011}. 
 \citet{Bouret2012} calculated the following stellar parameters: $T_{\rm eff} = 33.5 \pm 1.0$~kK and 
 $\log\,g ~=~3.42 \pm 0.1$. Similar sets of parameters were estimated 
by \citet[][$T_{\rm eff} = 33.5 \pm 0.5$~kK, $\log\,g~~=~3.71 \pm 0.05$, and $R_\star = 20.7 \pm 0.6$~R$_{\sun}$]{Hawcroft2021}
and \citet[][$T_{\rm eff} = 34$~kK, $\log\,g = 3.25$, $R_\star = 19.8$~R$_{\sun}$, and $v~\sin~i~=~100$~km~s$^{-1}$]{Gormaz2022}.
A projected rotation velocity of $v\sin~i = 82$~km~s$^{-1}$, close to the previous one, was measured by \citet{Holgado2022}. These stellar parameters lead to a rotation period of less than $12.5$~d.
The Fourier analysis of the time series of spectroscopic observations made by \citet{Rauw2001} exhibits recurrent variability in the absorption components of the P-Cygni profiles of \ion{He}{ii}~$\lambda$4686 \AA\, and H$\alpha$ with a period of roughly $4.8$~d long.

\begin{figure}
    \centering
    \includegraphics[width=0.9\linewidth]{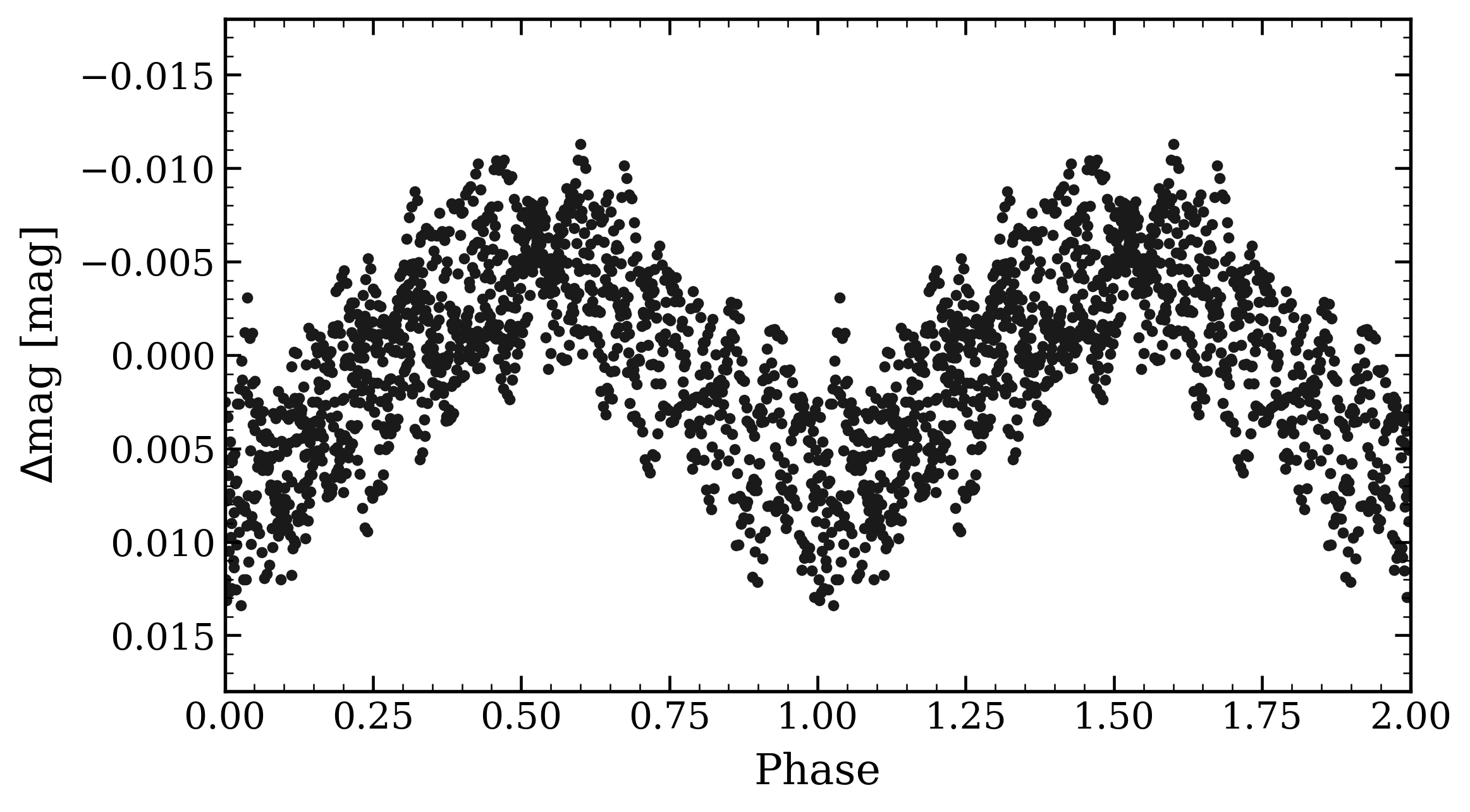}\\
    \includegraphics[width=0.9\linewidth]{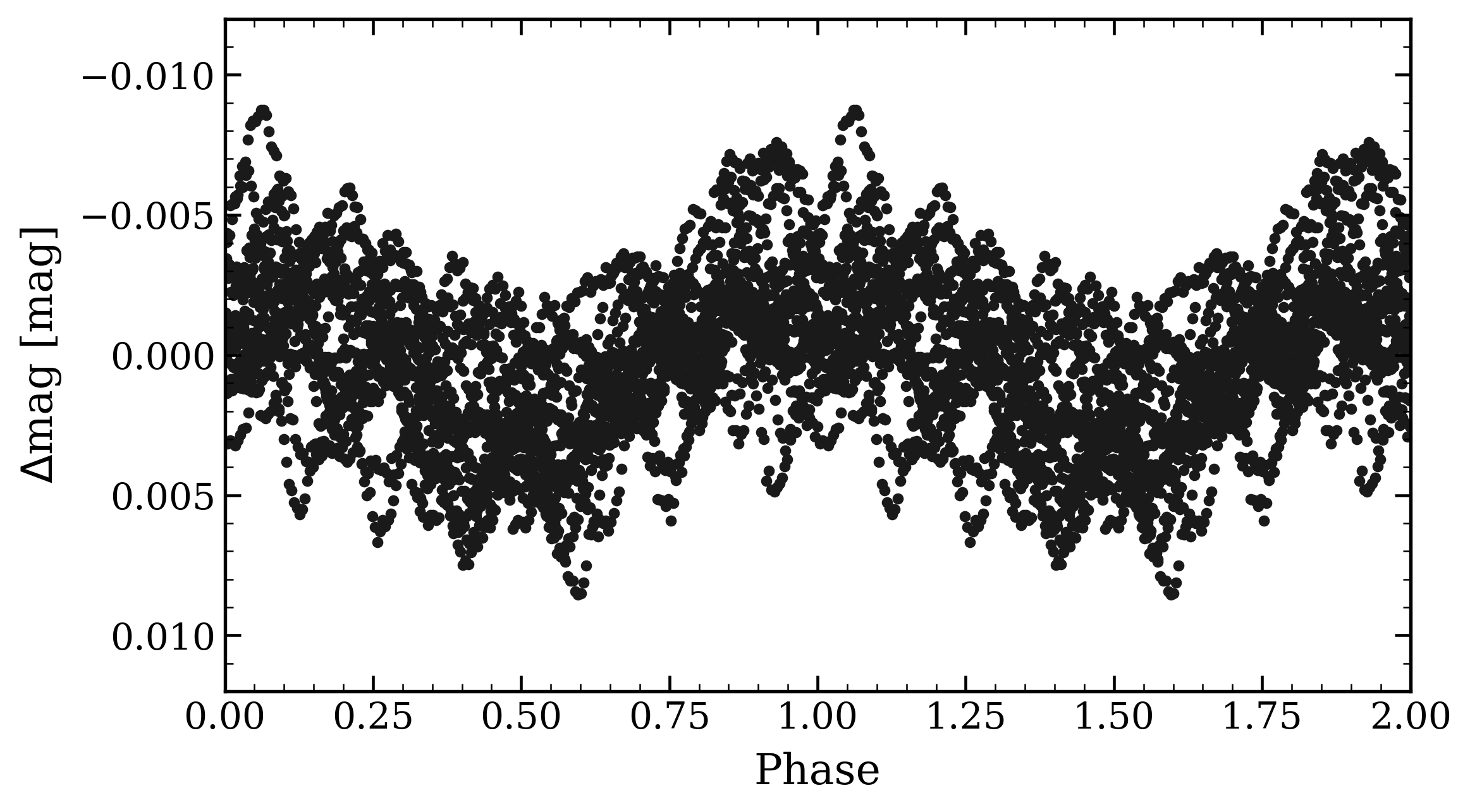}
    \caption{Phase-folded diagram of \object{HD~192639} obtained from sector $41$ of TESS light curves using periods of $4.833$~d {\it(upper panel)} and $1.55$~d. {\it (bottom panel)}.}
    \label{fig:PD_HD192639}
\end{figure}

\begin{figure}
    \centering
\includegraphics[width=0.8\linewidth]{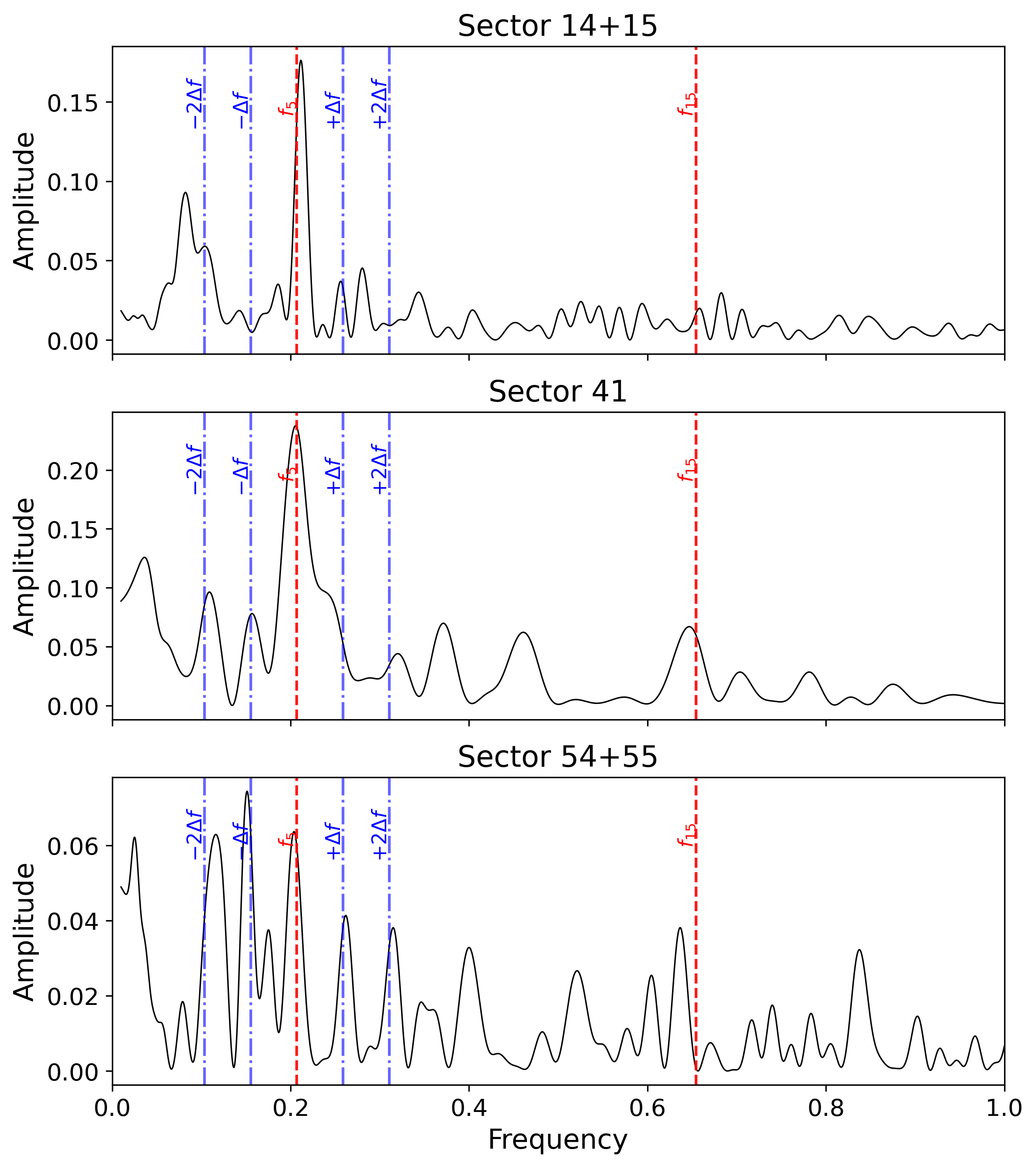}
\caption{Periodogram highlighting a series of equally spaced frequency peaks in HD 192639. The vertical dashed lines mark the positions of the expected frequencies with an equidistant spacing of $\Delta f=0.05185$~d~$^{-1}$. Several frequency peaks also appear at integer multiples of this spacing.}
\label{HD192639_periodogram}
\end{figure}

The light curves of \object{HD~192639} were obtained from TESS observations across sectors $14, 15, 41, 54,$ and $55$ (see Fig.~\ref{fig_lightcurveHD192639}). 
To enhance the resolution and precision in detecting low frequencies, we combined the light curves from consecutive sectors. The list of the most significant frequencies is given in Table~\ref{table:HD192639}.
Among them, we identify $f_5~=~0.2071$~d$^{-1}$, corresponding to a period of $4.83$~d, which exhibits the highest power and is responsible for the spectroscopic variations reported by \citet{Rauw2001}. This frequency, identified via the pulsation model as a $g$-mode pulsation (either $\ell=1$ or $\ell=2$), produces a coherent phase-folded signal in the light curve, as seen from TESS sector $41$ in Fig.~\ref{fig:PD_HD192639}. We also find that the observed data folded with a period of $1.55$~d display a coherent modulation, as shown in Fig.~\ref{fig:PD_HD192639} (bottom panel). We identify this oscillation in Table~\ref{HD192639_analysis} as $f_{15}$. 

The analysis of the average frequency spacing listed in Table~\ref{HD192639_analysis} reveals a rotational splitting centred on $f_5$. Using the mean-square method, we obtain an average frequency separation of $\Delta f=0.05185$~d$^{-1}$. This spacing also appears as a nearly constant separation among several other detected frequencies (see Fig.~\ref{HD192639_periodogram}). The average separation $\Delta f$ may be interpreted as rotational splitting, and the presence of a quintuplet pattern in the observed frequency spectrum is consistent with a $g$ mode with $\ell=2$. At the same time and intriguingly, all the observed frequencies appear to be organised as integers or half-integer multiples of the derived frequency separation, which is also indicative of an underlying rotational modulation. Such a scenario is expected when the excited frequency satisfies $f_0 << f_{\rm rot}$ \citep{Saio2018} or $f_0 \simeq f_{\rm rot}$, being $f_{\rm rot}$ the rotation frequency\footnote{For a mode with azimuthal order $m$, the relation between a frequency measured in the inertial frame ($f$) and in the corotating frame ($f_0$) is $f~=~f_0+~m~f_{\rm rot}$}. The latter would occur for variability induced by a surface inhomogeneity (e.g. a long-lived spot) co-rotating with the star. This condition is also expected for high-$\ell$ modes, whose $r$-mode frequencies are of the order of the rotation frequency ($m~ \Omega$). Figure~\ref{HD192639_periodogram} displays
many equidistant peaks and multiples of $\Delta\,f$. 

\begin{figure}
    \centering
\includegraphics[width=0.8\linewidth]{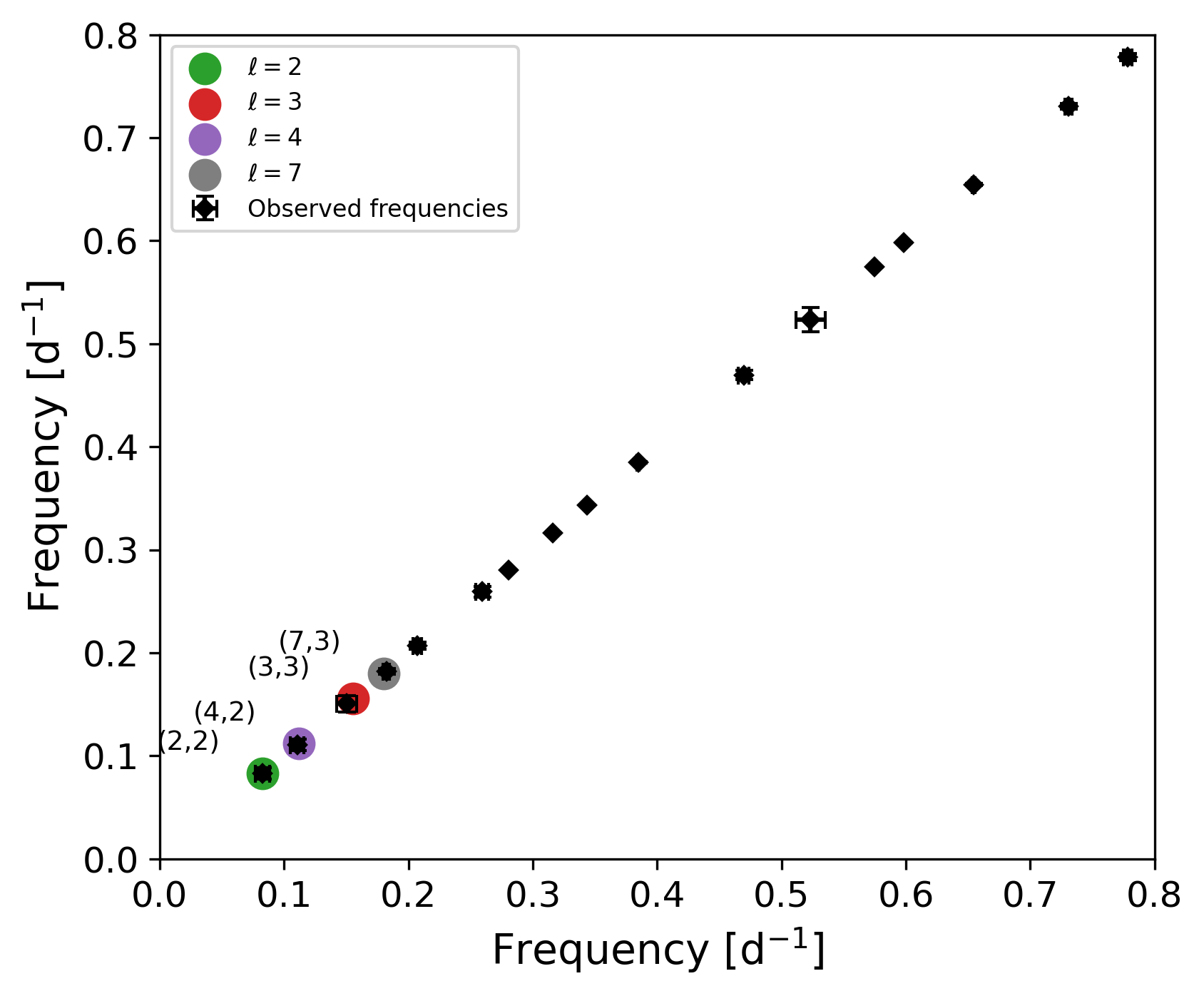}
    \caption{\object{HD 192639}: $r$ modes calculated with the rotation frequency, $f_{\rm rot}~=~0.0622$~d$^{-1}$ using several values of $(\ell,m)$. Average observed frequencies, along with the standard deviation, are indicated by a black symbol.}
\label{fig:192639_Rossby}
\end{figure}

In Table~\ref{HD192639_analysis}, we also identify the $g$ modes computed with {\tt LPCODE} for a $\sim 40~\mathrm{M}_\odot$ model (which fits the luminosity derived from Table~\ref{theoretical_comparison}; see Fig. \ref{fig:HR-LPCODE}). Moreover, from the particular frequency spacing found among the observed frequencies, we interpret these as a result of the rotational splitting (see discussion in Sect.\ref{Discussion}). Then, we applied the correction given in Eq.~(\ref{eq:gmode}) and obtained $f_{\rm rot}=0.0622$~d$^{-1}$.   The rotation frequency yields a rotation period of $16.1\pm2.6$~d. We then computed the Rossby modes, shown in Fig.~\ref{fig:192639_Rossby}. These candidate $r$ modes are identified by their $(\ell,m)$ indices in Table~\ref{HD192639_analysis}.
 
\subsection{Analysis of HD~195592}

\object{HD~195592} is a massive runaway star \citep{McSwain2007} showing $\gamma$-ray emission produced by the inverse Compton up-scattering of photons from heated dust present in the bow-shock structure \citep{delValle2013}. It was classified as an O9.7~Ia star by \citet{Sota2011}. The star has been proposed several times as a possible SB1 binary \citep{McSwain2007} or triple system \citep{Becker2010, Gormaz2022}. Using high-quality spectroscopic observations in the blue domain, \citet{Becker2010} found two independent periods of variability for this star ($5.063$~d and $\sim 20$~d) and constrained the rotation period to the interval [2.7, 19.4] days. 

\begin{figure}
    \centering    \includegraphics[width=0.9\linewidth]{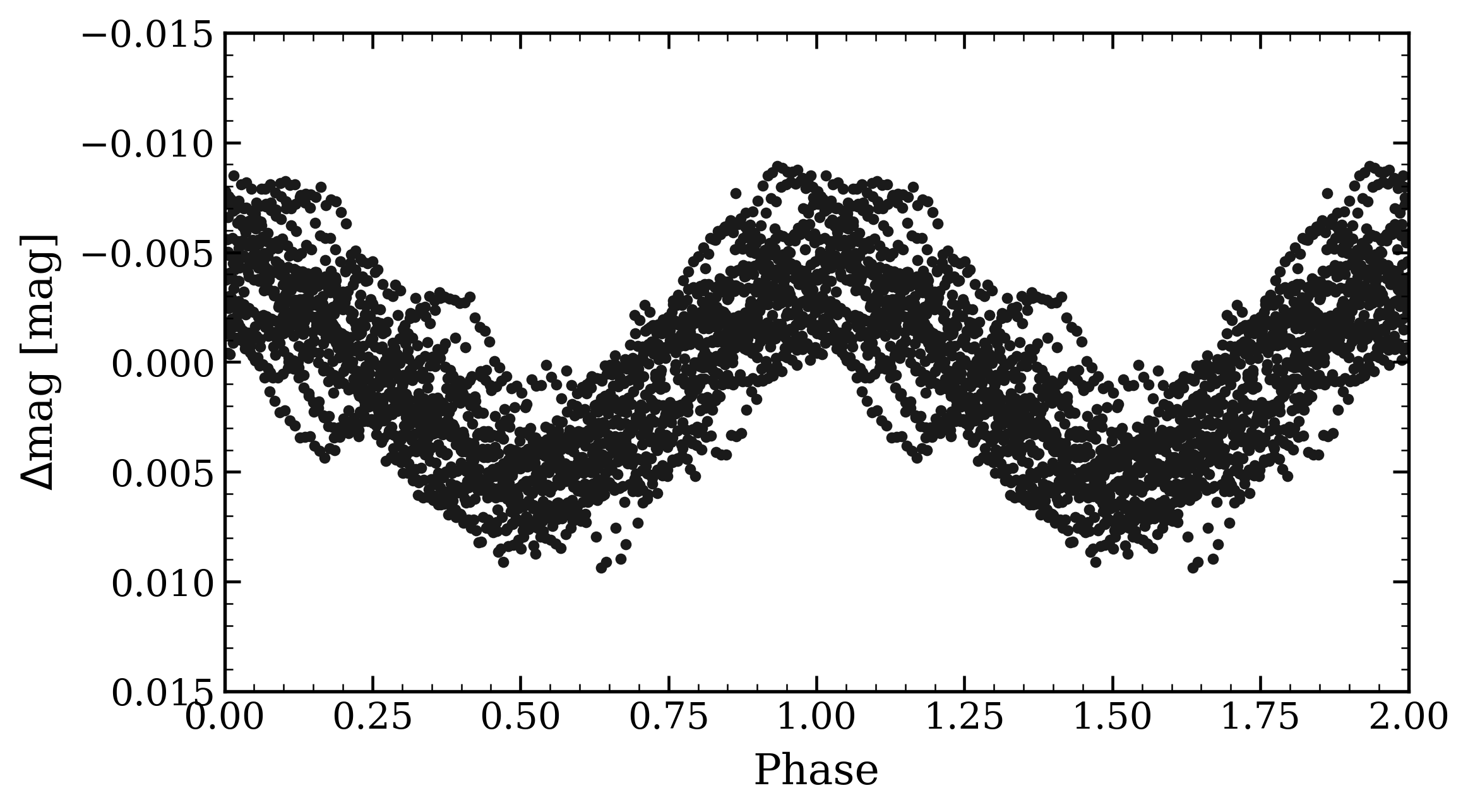}
    \caption{Phase-folded diagram for \object{HD~195592} obtained with TESS light curves from sector $14$ and $15$. A coherent signal arises using $P~=~2.45$~d ($f=0.4084$~d$^{-1}$).}
    \label{fig:PD_HD195592}
\end{figure}

\begin{figure}
    \centering
    \includegraphics[width=0.8\linewidth]{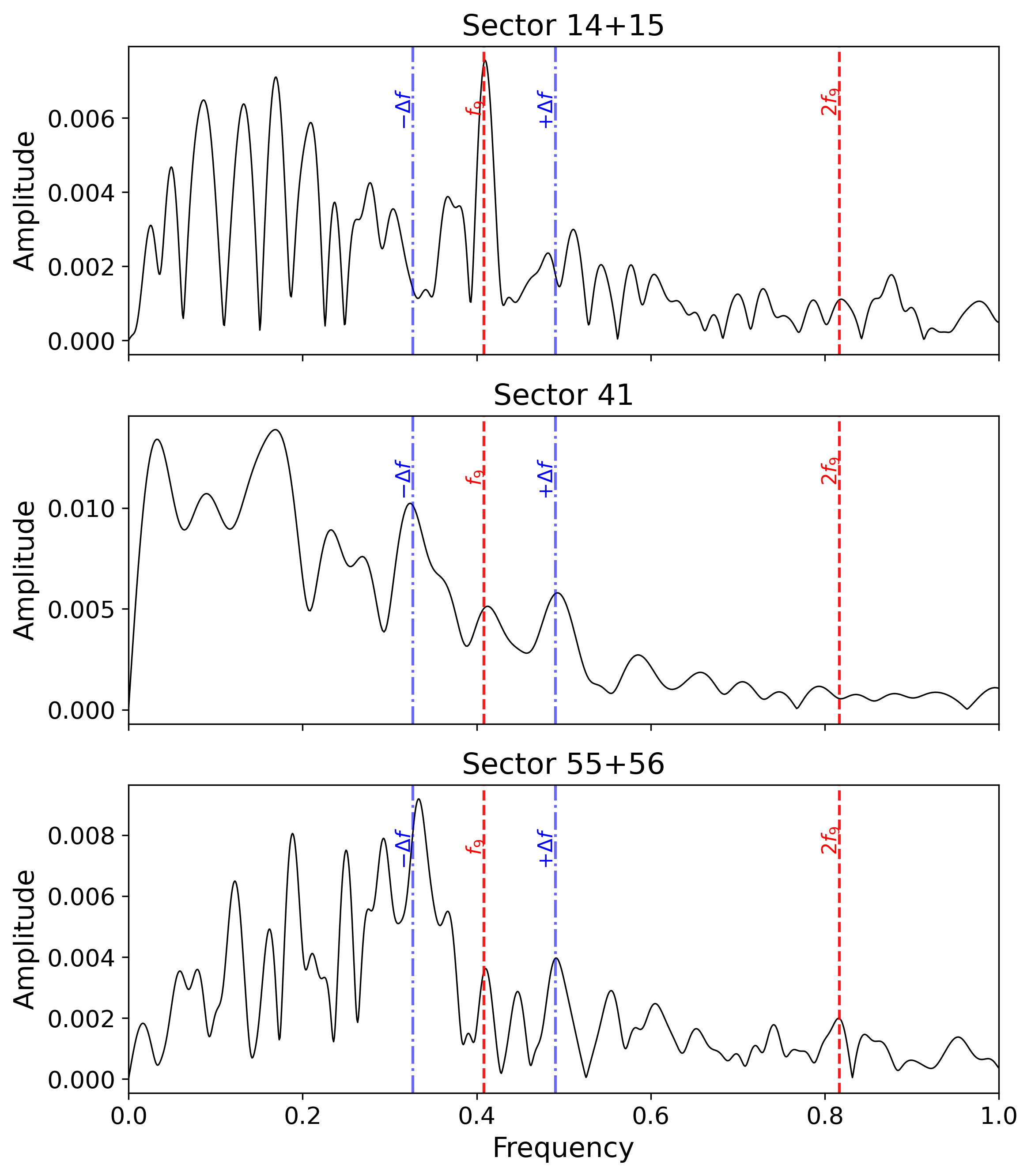}
\caption{Periodograms for HD 195592 highlighting a series of equally spaced frequency peaks. The vertical dashed lines mark the positions of the
    expected frequencies with an equidistant spacing of $\Delta f=0.08185$~d$^{-1}$ centred on $f_9$. Several frequency peaks also appear at integer multiples of this spacing.}
\label{fig:periodogram_HD195592}
\end{figure}

\begin{figure}
    \centering
    \includegraphics[width=0.8\linewidth]{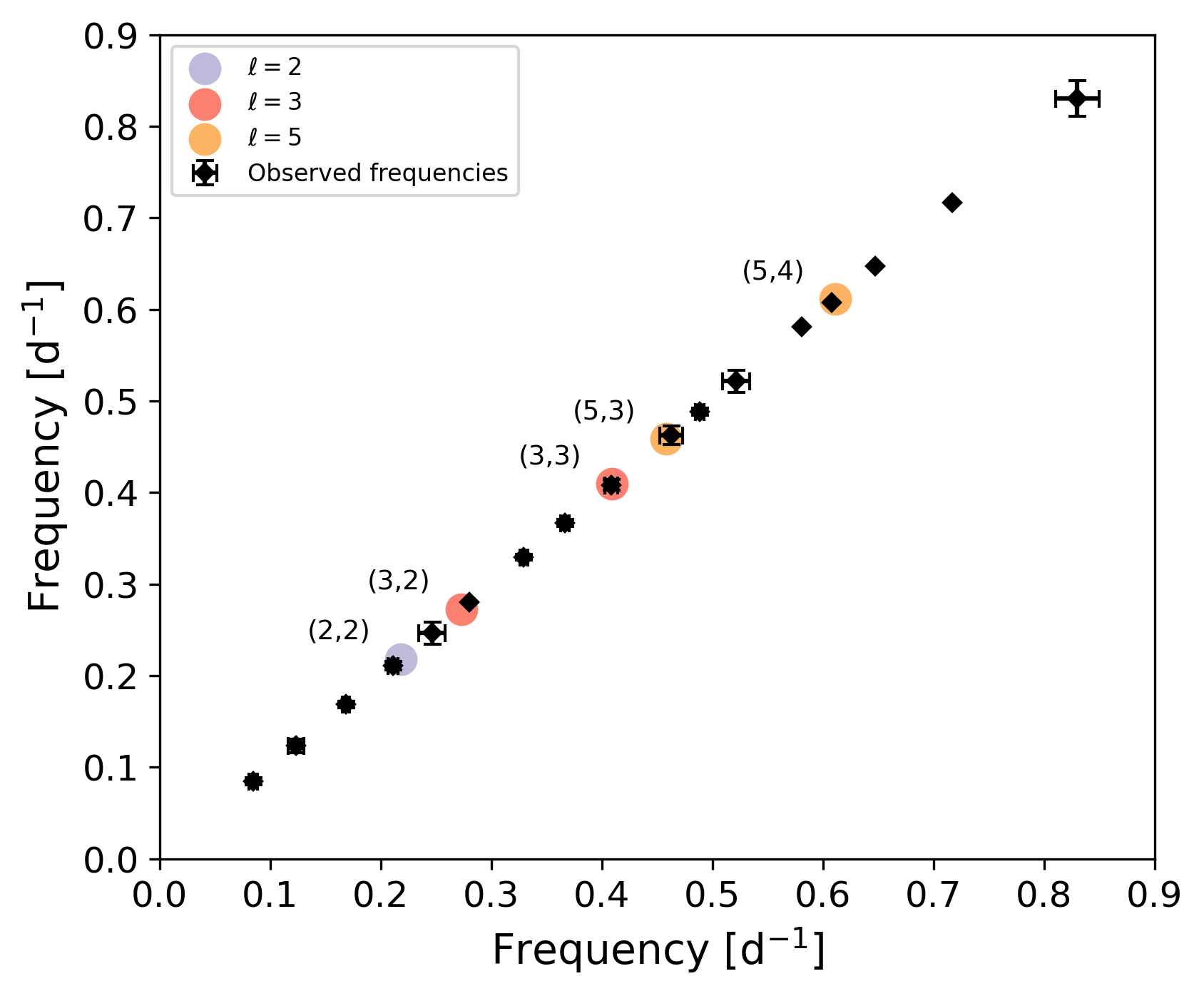}
\caption{\object{HD~195592}: $r$ modes calculated using the rotation frequency, $f_{\rm rot}=0.1637$ d$^{-1}$, and several values of $(\ell,m)$.  Average observed frequencies, along with their standard deviations, are indicated by black symbols.}
\label{fig:195592_Rossby}
\end{figure}

In Fig.~\ref{fig:lightcurveHD195592}, we present the TESS light curves from sectors $14$ and $15$, $41$, and $55$ and $56$ of \object{HD\,195592}, together with their corresponding scalograms and periodograms. We detect $26$ significant frequencies in the TESS observations, listed in Table~\ref{table:HD195592}. The scalograms reveal that the power of several frequencies undergoes temporal modulation, increasing and decreasing across the different sectors, whereas the frequencies
$f_{3}=0.0847$~d$^{-1}$ ($11.80$~d)  and $f_{7}=0.1685$~d$^{-1}$ ($5.93$~d) remain almost constant across the sectors. 
 Among the listed frequencies, we find an average period of $19.16$~d, which is close to the $20$~d value reported by \citet{Becker2010}. We also identify two other periods, one with a period of $5.34$~d with high power and another with a period of $4.74$~d with lower power (see Table~\ref{table:HD195592}). These values are very close to the previously reported value of $5.06$~d.  In addition, we find that the frequency $f_9=0.4084$~d$^{-1}$ ($P=2.45$~d) produces a clear but weak periodic signal (of about $10$~mmag) in the TESS light curve, as illustrated in the phase-folded diagram in Fig.~\ref{fig:PD_HD195592}. This frequency can be identified as a possible $g$-mode oscillation ($\ell=1$), as predicted by the pulsation model of a $30$~M$_\sun$ star (see Table~\ref{theoretical_comparison}).
 Table~\ref{analysis_HD195592} summarises the selected frequencies and their corresponding frequency separations, both between consecutive orders and among the highest peaks. As we combined two consecutive sectors, the resulting time span yields a more restrictive Rayleigh frequency resolution of $1.5/\Delta T=0.028$~d$^{-1}$.
 We identify several frequencies with approximately constant spacing, highlighted in Table~\ref{analysis_HD195592} with boldface fonts (as a reference, we selected the frequency at $f_9$).  A mean-square optimisation yields a best-fitting separation of $\Delta f=0.08185$~d$^{-1}$ (illustrated in the periodograms of Fig.~\ref{fig:periodogram_HD195592}). We interpret this constant spacing as a rotational splitting, where it is possible to identify, at least, a triplet around $f_9$. Interestingly, most of the frequencies listed in Table \ref{analysis_HD195592} also appear to be either integer or half-integer multiples of $\Delta f$, suggesting that a significant fraction of the detected variability is modulated by rotation. Moreover, we find that many of the observed frequencies match the $g$-mode oscillations computed in Table \ref{theoretical_comparison} for a $30$~M$_\sun$ star (see also Fig. \ref{fig:HR-LPCODE}). Then, using the first-order rotational splitting, Eq.~(\ref{eq:gmode}), for a dipole configuration (where $C(\ell,m)=0.5$), we derive a rotational frequency of $f_{\rm rot}=0.1637$~d$^{-1}$. This leads to a rotation period of $6.1\pm1.0$~d. Adopting the derived value $f_{\rm rot}$, we searched for signatures of Rossby modes. Figure~\ref{fig:195592_Rossby} presents a comparison between the theoretically expected Rossby frequencies and those observed. The frequency pattern ($f= m~f_{\rm rot}$) is primarily consistent with high-$\ell$ Rossby modes, characterised by $(\ell,m) = (2,2), (3,2)$,  $(3,3)$, $(5,3)$, and $(5,4)$. 

\vskip 0.5cm

\section{Discussion}
\label{Discussion}
 
Using TESS photometric light curves, we analysed three late O-type supergiants that exhibit a characteristic frequency pattern with an approximately uniform frequency spacing. Moreover, from {\tt LPCODE} evolutionary models, we identified the corresponding $g$ modes, which, unexpectedly, do not show any regular period spacing, indicating that these oscillations lie outside the asymptotic regime. 

Since rotational effects on the pulsation frequencies are expected to be weak with $(\Omega/\Omega_g)^2 \ll 1$, we applied {\tt LPCODE} to non-rotating stellar models to compute the theoretical $g$-mode frequencies. Using the obtained rotation periods of $14.5\pm4.2$~d, $16.1\pm2.6$~d, and $6.1\pm1.0$~d for \object{HD~188001}, \object{HD~192639}, and \object{HD 195592}, respectively, we obtained $(\Omega/\Omega_g)^2$ values of approximately $1\%$, $1.6\%$, and $12\%$ (based on the results of Table~\ref{tab_X}). These values remain sufficiently low to justify neglecting rotational effects in the pulsation calculations.

One important result from the stellar evolution models is that the positions of the stars in the Hertzsprung–Russell (HR) diagram (shown in Fig.~\ref{fig:HR-LPCODE}), in terms of effective temperature and luminosity, indicate that they are near the end of the main-sequence phase, still undergoing core H-burning. This is not inconsistent with the O-type supergiant spectroscopic classification as supergiants, since luminosity class is not a direct diagnostic of the nuclear-burning stage \citep{Martins2017}. Indeed,  \citet{Bellinger2024} show that blue supergiants may correspond to different internal evolutionary states, including
core H-burning, H-shell-burning, and core He-burning configurations, and that asteroseismology can help distinguish between them.
In this context, the spectroscopic characteristics primarily reflect the presence of strong radiatively driven winds, in which radiation pressure significantly affects the structure and dynamics of their stellar atmospheres, producing the observational features typically associated with supergiants.

It is also important to emphasise the high degree of similarity among the frequencies identified in the observed oscillations of the three supergiants in which low-frequency $g$ modes are excited. These pulsational properties can be reproduced by stellar models in the late main-sequence phase, supporting the interpretation that these stars continue to undergo core H-burning.  
In addition, we inspected the periodogram of other related stellar objects, mainly the B-type supergiants reported by \citet{Kourniotis2025} and $\rho$~Leo, which has an extensive photometric time coverage from both K2 and TESS missions, together with a large set of spectroscopic observations \citep{Checha2025}. The frequency analysis of these studies also reveals signals in a similar frequency range, with values comparable to those found in our O-type supergiant sample, thereby providing independent support for the physical reality of the detected frequencies.

Another important observation is that our star sample exhibits clear signs of rotational modulation. The presence of integer multiplets in the spacing frequency reveals this phenomenon. 
However, rotation itself does not produce significant shifts in the frequencies of the detected $g$ modes, although it may give rise to rotational splitting, revealing multiplet structures in the observed frequency spectrum \citep[see e.g.][]{Kurtz2014}.

The presence of frequencies that are integer multiplets of the spacing frequency may arise when: i) the oscillation frequency   $f_0 << m f_{\rm rot}$, leading to a superperiod effect\footnote{The superperiod is defined as $m$ times the oscillation period, where $m$ is the azimuthal order.} \citep{Saio1990}; ii) $f_0=f_{\rm rot}$, in which case the variability can be naturally interpreted as arising from corotating features (e.g. magnetic structures, hot spots, or surface inhomogeneities); or iii) high orders of $r$ modes are excited ($\sigma \simeq m \Omega$)  \citep{Papaloizou1978}. From an observational perspective, case iii) appears to be the most plausible scenario, as it naturally explains the observed frequency splitting and the presence of RWs. However, a detailed theoretical treatment of the excitation and development of $r$ modes in rotating stars is beyond the observational scope of the present work and is therefore left for future studies. 

Classical RWs were first predicted to exist in the Sun and stars over forty years ago \citep[cf.][]{Zaqarashvili2021}. Given that O-type supergiants typically have non-negligible rotation velocities, the Coriolis effect is expected to play a significant role in their internal fluid dynamics, leading to the formation of RWs and IGWs. The few observational detections of inertial gravity waves in the literature have been for massive stars of spectral type O and were made by matching the observed morphology of the low-frequency power excess with that predicted by state-of-the-art simulations of IGWs \citep{Rogers2013, Rogers2015, Aerts2015, SimonDiaz2017, Aerts2018}.

One of the main advantages of detecting rotational splitting is that they provide direct constraints on the stellar rotation properties. From the measured splitting, we can derive the rotation period, the rotational velocity, and (if the stellar radius is known) the inclination angle of the rotation axis. 
The derived quantities are summarised in Table~\ref{tab_X}, which lists the stellar radius, the rotation period,  the derived rotational velocity ($2\,\pi\, R_\star/P_{\rm rot})$, the inclination angle of the rotation axis ($i$) from $v\,\sin\,i$, the critical rotation velocity, $v_{c}= (\frac{2}{3}\, G\, M_\star/R_\star)^{1/2}$ \citep[see][]{Maeder2000}, and the linear rotation rate, $\tilde{\Omega}=v_{\rm rot}/v_{\rm c}$. We used the initial mass of the evolutionary track to derive $v_{\rm c}$. Uncertainties in the values are indicated.
Furthermore, we assumed that the difference between the surface and core rotation is moderate. Only small frequency shifts were observed when comparing the measured frequencies with the theoretical values from the dispersion relation (Eq.~(\ref{Inertial frame})). A rotation profile consistent with uniform rotation for the $\sim 8$~M$_\sun$ star
\object{HD~157056} was also reported by \citet{Briquet2007}.

Our results indicate that \object{HD~188001} and \object{HD~192639} are likely seen close to equator-on, while \object{HD~195592} may be observed at an inclination of $i\sim 20^\circ$. 
Moreover, the observed frequency patterns are compatible with both global RWs (mainly, $(2,2)$, $(3,2)$, or $(3,3)$ modes) and the identified $g$ modes. The overlap between the expected frequency domains of these oscillations prevents a unique mode identification. 
Furthermore, it is noteworthy that these stars are classified as runaways. Their runaway nature may have affected their rotational properties and internal structure, likely as a result of past dynamical interactions or binary evolution. 

\begin{table}[h]
\centering
\caption{Rotation period and inclination angles of the rotation axis obtained from the pulsation analysis. The critical rotation velocity and $\tilde{\Omega}$ are given as a reference.}
\tabcolsep 2.3pt
\begin{tabular}
{@{}rrrrccc@{}}
\hline
\hline
HD~~& $R_\star$~ & $P_{\rm rot}$~ & $v_{\rm rot}$ ~~~& $i$ &  $v_c$ & $\tilde{\Omega}$\\ 
& [R$_{\odot}$] &  [d]~ & [km~s$^{-1}$] &  [$^\circ$] &[km~s$^{-1}$] & \\ 
\hline
\rule{0pt}{1em}%
$188001$ & $23.0\pm1$ & $14.5\pm4.2$ & $80\pm23$ & $\sim 90$  & $526$ & $0.15$\\
$192639$ &  $19.8\pm1$ & $16.1\pm2.6$ & $62\pm11$  & $\sim 90$ &$507$ &$0.12$\\
$195592$ &   $21.5\pm1$ & $6.1\pm1.0$ & $178\pm32$  & $20\pm5$  &$421$ & $0.42$\\
\hline
\end{tabular}
\label{tab_X}
\end{table}

The observed rotational modulation and Rossby oscillations appear to be recurrent or long-lived, as they are observed along different sectors that span approximately five years. Over this period, they are detected with varying amplitudes and show slight frequency shifts across different observing sectors.
 These waves might also contribute to frequency fluctuations characterised by a power excess at low frequencies in the periodogram, known as red noise \citep{Ramiaramanantsoa2018, Bowman2019, Thompson2024}. Additional research is required to evaluate the impact of RWs on the spectrograms of massive blue stars. 

\section{Conclusions}
\label{Conclusions}

We analysed TESS light curves of three late O-type supergiant stars and identified stable $g$-mode oscillations and signatures consistent with rotational modulation. We also emphasised the remarkable similarity in the frequency values and their particular pattern $\sigma= m~\Omega$ across two stars (\object{HD~192639} and \object{HD~195592}). Comparable frequencies in the same low-frequency domain have also been reported in other B-type supergiants, suggesting that these oscillations may be a common property of evolved massive stars. Stellar evolution and pulsation models predict $g$-mode oscillations near the end of the main-sequence phase, while the supergiant characteristics inferred from spectroscopy mainly reflect atmospheric properties associated with strong radiatively driven winds. In this sense, the pulsation behaviour provides a more direct probe of the internal stellar structure, whereas the supergiant classification is primarily linked to atmospheric and wind properties.

Our results also provide tentative evidence that RWs may be excited in these objects, suggesting that such large-scale inertial oscillations could contribute to the observed low-frequency variability. This variability could also be contributing to the red noise phenomenon. A longer time baseline, and hence improved frequency resolution, is required to characterise the frequency patterns more robustly and to constrain the stellar rotation rates and the distribution of inclination angles better. In addition, a dedicated theoretical investigation of the $r$ modes is essential to explain the observed frequency patterns and probe the internal rotation profile of these stars.

   \begin{acknowledgements}
   Some/all of the data presented in this paper were obtained from the Mikulski Archive for Space Telescopes (MAST). STScI is operated by the Association of Universities for Research in Astronomy, Inc., under NASA contract NAS5-26555. The NASA Office of Space Science provides support for MAST for non-HST data via grant NNX13AC07G and by other grants and contracts.
      This project received funding from the European
Union’s Framework Programme for Research and Innovation Horizon 2020
(2014-2020) under the Marie Sk{\l}odowska-Curie Grant Agreement No. 823734 - POEMS. It has also been co-funded by the European Union, Project 101183150 - OCEANS. AA and LC thank the financial support from CONICET (PIP 1337) and the Universidad Nacional de La Plata (Programa de Incentivos 11/G160), Argentina. AC and GA acknowledge support from Centro de Estudios Atmosféricos y Astroestadística (CEAAS), Universidad de Valparaíso, Chile. AL acknowledges in part funding by the Belgian Federal Science Policy Office - Policy for Science Program Contract No. P4S/251/Gaia-BRASS.
SG, MK, and JSA acknowledge financial support from the Czech Science Foundation (GA \v{C}R, grant number 25-17532S). The Astronomical Institute of the Czech Academy of Sciences is supported by the project RVO:67985815. AHC thanks Marcelo Miller Bertolami for his assistance in handling {\tt LPCODE}.
\end{acknowledgements}

\bibliographystyle{aa} 
\bibliography{reference} 

\begin{appendix}
\onecolumn
\section{$g$ modes computed with {\tt LPCODE} evolutionary stellar models}

\begin{table*}[h!]
\centering
\caption{$g$-mode oscillations ($\ell=1$ and $\ell=2$) for models with initial masses of $30$, $40$, and $50$~M$_\sun$, representative of the stars \object{HD~195592}, \object{HD~192639}, and \object{HD~188001}.}
\label{theoretical_comparison}
\begin{tabular}{cccc|ccc|ccc}
\hline
\hline
\noalign{\vskip 2pt}
& \multicolumn{3}{c|}{\object{HD 195592} ~~($30~{\rm M}_\sun$)} & \multicolumn{3}{c|}{\object{HD 192639}~~($40~{\rm M}_\sun$)}& \multicolumn{3}{c}{\object{HD 188001}~~($50~{\rm M}_\sun$)}\\
$\ell=1$ & $f$& $P$ & ~$\Delta P$ & $f$ & $P$ & ~$\Delta P$ &  $f$ & $P$& ~$\Delta P$ \\
~$k$&[d$^{-1}$]& [d]&[d] &[d$^{-1}$]&[d] & [d]& [d$^{-1}$]& [d] & [d]\\
\hline
\noalign{\vskip 2pt}
1 & $1.7844$ & $0.560$ & $0.059$ & 
$2.1943$&$0.456$ & $0.020$& 
$1.7325$&$0.577$ & $0.072$\\
2& $1.6751$ & $0.597$ & $0.037$ & $1.4783$&$0.676$ & $0.221$ &
$1.2010$&$0.833$ & $0.255$\\
3 & $1.2445$ & $0.804$ & $0.207$ & $0.9203$&$1.087$ & $0.410$ & 
$0.8024$&$1.246$ & $0.414$ \\
4 & $0.8318$ & $1.202$ & $0.399$ & $0.5994$&$1.668$ & $0.582$ & 
$0.4749$&$2.106$ & $0.859$\\
5 & $0.5936$& $1.685$ & $0.482$ & $0.5398$&$1.853$ & $0.184$ & 
$0.4194$&$2.384$ & $0.279$\\
6 & $0.5430$&$1.842$ & $0.157$ & $0.4422$&$2.261$ & $0.409$ & 
$0.3686$&$2.713$ & $0.329$\\
7 & $0.5181$ &$1.930$ & $0.088$ & $0.3574$&$2.798$ & $0.537$ & 
$0.3207$&$3.118$ & $0.405$\\
8 & $0.3969$ &$2.519$ & $0.589$ & $0.3518$&$2.843$ & $0.044$ & 
$0.2445$&$4.090$ & $0.971$\\
9 & $0.3742$&$2.672$ & $0.153$ & $0.2847$&$3.513$ & $0.670$ & 
$0.2431$&$4.113$ & $0.023$\\
10 & $0.3454$&$2.895$ & $0.223$ & $0.2711$&$3.689$ & $0.177$ & 
$0.2110$&$4.738$ & $0.625$\\
11 & $0.3024$&$3.307$ & $0.411$ & $0.2571$&$3.890$ & $0.201$ &
$0.1896$&$5.273$ & $0.535$\\
12 & $0.2748$ &$3.639$ & $0.333$ & $0.2282$&$4.383$ & $0.493$ & 
$0.1716$&$5.828$ & $0.555$\\
13 & $0.2558$ &$3.909$ & $0.270$ & $0.2049$&$4.879$ & $0.497$ & 
$0.1507$&$6.634$ & $0.807$\\
14 & $0.2130$&$4.694$ & $0.785$ & $0.1972$&$5.071$ & $0.191$ & 
$0.1383$&$7.232$ & $0.597$\\
15 & $0.2126$& $4.704$ & $0.010$ & $0.1771$&$5.646$ & $0.575$ & 
$\cdots$&$\cdots$& $\cdots$\\
16 & $0.1759$& $5.684$ & $0.980$ & $0.1602$&$6.244$ & $0.598$ & 
$\cdots$&$\cdots$&$\cdots$\\
17 & $0.1694$& $5.904$ & $0.220$ & $0.1591$&$6.287$& $0.043$&
$\cdots$ & $\cdots$ & $\cdots$\\
18 & $0.1680$& $5.954$ & $0.050$ & $\cdots$& $\cdots$ &$\cdots$ &
$\cdots$& $\cdots$ &$\cdots$\\
19 & $0.1633$ & $6.125$ & $0.171$ &  $\cdots$&$\cdots$ &$\cdots$&
$\cdots$& $\cdots$ &$\cdots$\\
20 & $0.1532$& $6.529$ & $0.404$& 
$\cdots$&  $\cdots$ &$\cdots$&
$\cdots$& $\cdots$ &$\cdots$\\
\hline
 $\ell=2$ & $f$& $P$ & ~$\Delta P$ & $f$ & $P$ & ~$\Delta P$ &  $f$ & $P$& ~$\Delta P$ \\
~$k$&[d$^{-1}$]& [d]&[d] &[d$^{-1}$]&[d] & [d]& [d$^{-1}$]& [d] & [d]\\
\hline
$1$ & $2.5539$ & $0.392$ &$0.056$ &$2.5513$ &$0.392$ &$0.075$ & $2.1622$ & $0.462$ & $0.112$\\
$2$ & $2.1204$ &$0.472$ &$0.080$ & $2.3339$&$$0.428$$ & $0.037$ & $1.9962$ & $0.501$ & $0.038$\\
$3$ & $1.9255$ &$0.519$ &$0.048$ &$2.2284$ &$0.449$ &$0.020$ & $1.7504$ & $0.571$ & $0.070$\\
$4$ & $1.7625$ &$0.567$ &$0.048$ &$1.5679$ &$0.638$ &$0.189$ & $1.3657$ & $0.732$ & $0.161$\\
$5$ &$1.4128$ &$0.708$ & $0.140$& $1.0215$ &$0.979$ &$0.341$ &$0.7682$ & $1.302$ & $0.570$\\
$6$ & $0.9692$& $1.032$& $0.324$&$0.8832$ &$1.132$&$0.153$ & $0.7217$ & $1.386$ & $0.084$\\
$7$ & $0.9241$ &$1.082$ &$0.050$ &$0.7623$ &$1.312$ &$0.180$ & $0.6363$ & $1.572$ & $0.186$\\
$8$ & $0.8898$&$1.124$ & $0.042$&$0.6068$ &$1.648$&$0.336$ & $0.5340$ & $1.873$ & $0.301$\\
$9$ &$0.6773$ &$1.476$ & $0.353$&$0.5972$ &$1.675$ &$0.027$ &  $0.4221$ & $2.369$ & $0.496$\\
$10$ & $0.6313$ & $1.584$& $0.108$&$0.4920$ &$2.033$ &$0.358$ & $0.4106$ & $2.435$ & $0.066$\\
$11$ &$0.5947$ & $1.682$& $0.097$&$0.4682$ &$2.136$ &$0.103$ & $0.3647$ & $2.742$ & $0.307$\\
$12$ & $0.5221$&$1.915$ &$0.234$ &$0.4360$ &$2.294$ &$0.158$ & $0.3236$ & $3.090$ & $0.348$\\
$13$ & $0.4639$&$2.156$ &$0.240$ &$0.3948$ &$2.533$ & $0.239$& $0.2965$ & $3.373$ & $0.283$\\
$14$ & $0.4424$&$2.260$ &$0.105$ &$0.3538$ &$2.827$ &$0.294$ & $0.2585$ & $3.869$ & $0.496$\\
$15$ & $0.3680$& $2.717$&$0.457$ &$0.3370$ &$2.967$ &$0.140$ & $0.2393$ & $4.179$ & $0.311$\\
$16$ & $0.3617$&$2.764$ &$0.047$ &$0.3060$ &$3.268$ & $0.300$& $0.2302$ & $4.344$ & $0.164$\\
$17$ & $0.3018$&$3.314$ & $0.549$& $0.2768$&$3.613$ &$0.346$ & $0.2258$ & $4.430$ & $0.086$\\
$18$ & $0.2919$&$3.426$ &$0.112$ &$0.2748$ &$3.639$ &$0.026$ & $0.2174$ & $4.600$ & $0.170$\\
$19$ & $0.2906$ &$3.441$ &$0.015$ &$0.2724$ &$3.671$ &$0.031$ & $0.2088$ & $4.789$ & $0.190$\\
$20$ & $0.2818$&$3.549$ &$0.108$ &$0.2422$ &$4.129$ &$0.458$ & $0.1889$ & $5.294$ & $0.504$\\
$21$ & $0.2649$&$3.775$ & $0.225$& $0.2236$&$4.471$ &$0.342$ & $0.1792$ & $5.579$ & $0.285$\\
$22$ & $0.2615$ &$3.824$ &$0.049$ &$0.2072$ &$4.826$ &$0.354$ & $0.1766$ & $5.662$ & $0.083$\\
$23$ &  $\cdots$&$\cdots$ & $\cdots$ &$0.1989$ &$5.028$ &$0.202$ & $0.1677$ & $5.962$ & $0.299$\\
$24$ &$\cdots$ & $\cdots$& $\cdots$&$0.1925$ &$5.194$ & $0.166$& $0.1552$ & $6.444$ & $0.482$\\
$25$ &$\cdots$ &$\cdots$ &$\cdots$ &$0.1899$ &$5.266$ &$0.072$ & $0.1503$ & $6.653$ & $0.209$\\
$26$ &$\cdots$ &$\cdots$ &$\cdots$ &$0.1802$ &$5.548$ &$0.283$ & $\cdots$ &$\cdots$ &$\cdots$\\
$27$ &$\cdots$ &$\cdots$ &$\cdots$ &$0.1721$ &$5.810$ &$0.262$ & $\cdots$ &$\cdots$ &$\cdots$\\
\hline
\end{tabular}
\tablefoot{The frequency is $f=1/P$. $k$ (first column) is the radial order of the modes. The periods ($P$) are given in days, and the corresponding period spacing ($\Delta P$) is also indicated. The models are characterised by structure parameters $[M_{\star}/M_{\sun}, T_{\rm eff}/(\rm{K}), \log(L_{\star}/L_{\sun}), X_{\rm H}(centre)]$ of $[30,29\,484,5.45,0.12]$, $[40,34\,007,5.65,0.19]$ and $[50,34\,550,5.83,0.21]$.}
\end{table*}

\twocolumn
    \section{Light curves and frequency analysis}
    \label{Ap:A}
    
This appendix provides the TESS light curves of \object{HD~188001}, \object{HD~192639}, and \object{HD~195592}, along with their WWZ scalogram, WWZ average power, and LS periodogram. Each plot has a vertical coloured bar which indicates the normalised power scale of the wavelet coefficients. The vertical scales for the frequency are displayed on the right. The light curves vary from sector to sector. Therefore, the average power at each frequency also changes between sectors, which is clearly visible in the scalograms.

\begin{figure}[h]
 \centering
\includegraphics[width=\columnwidth]{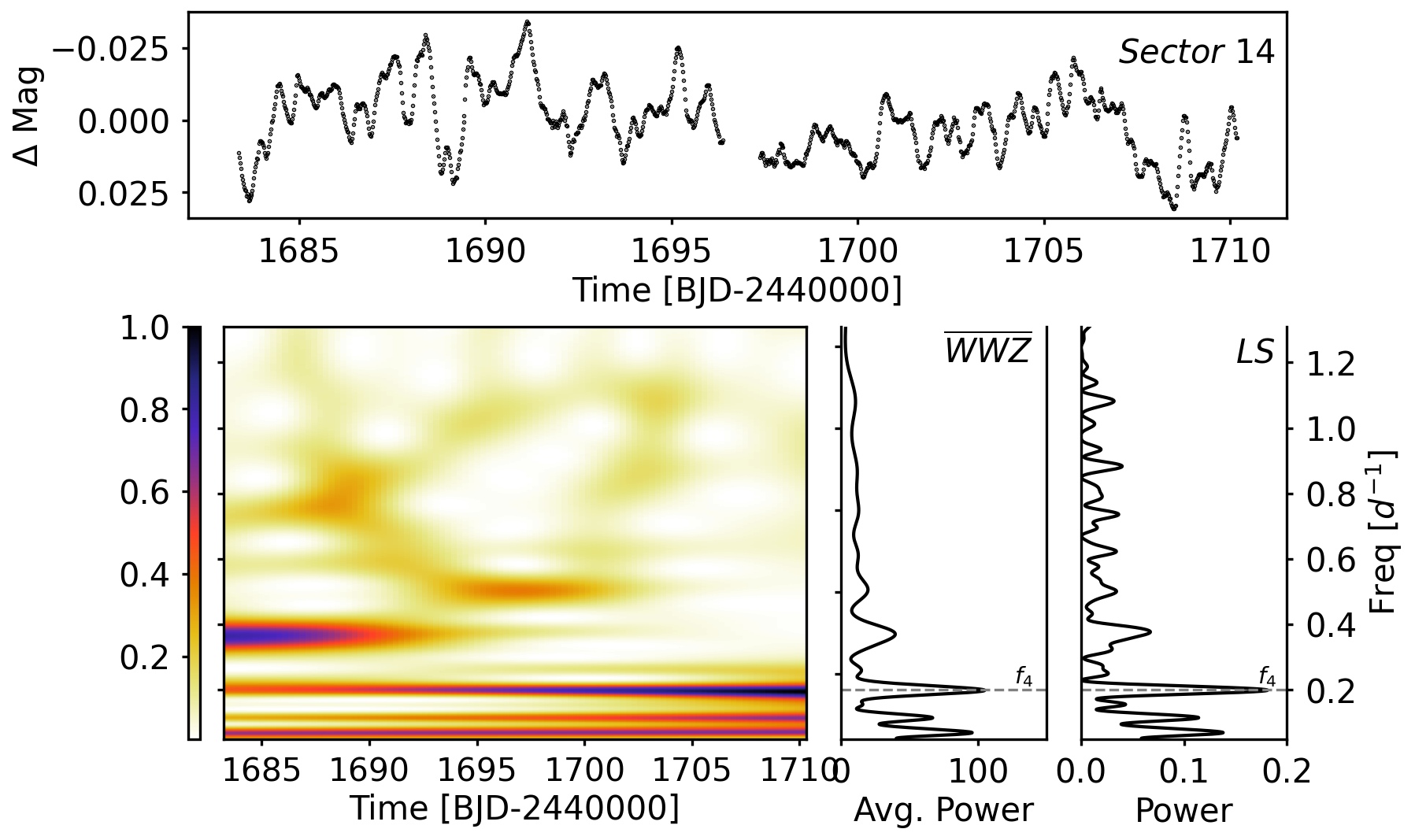}\  
\vspace{10pt}    \includegraphics[width=\columnwidth] {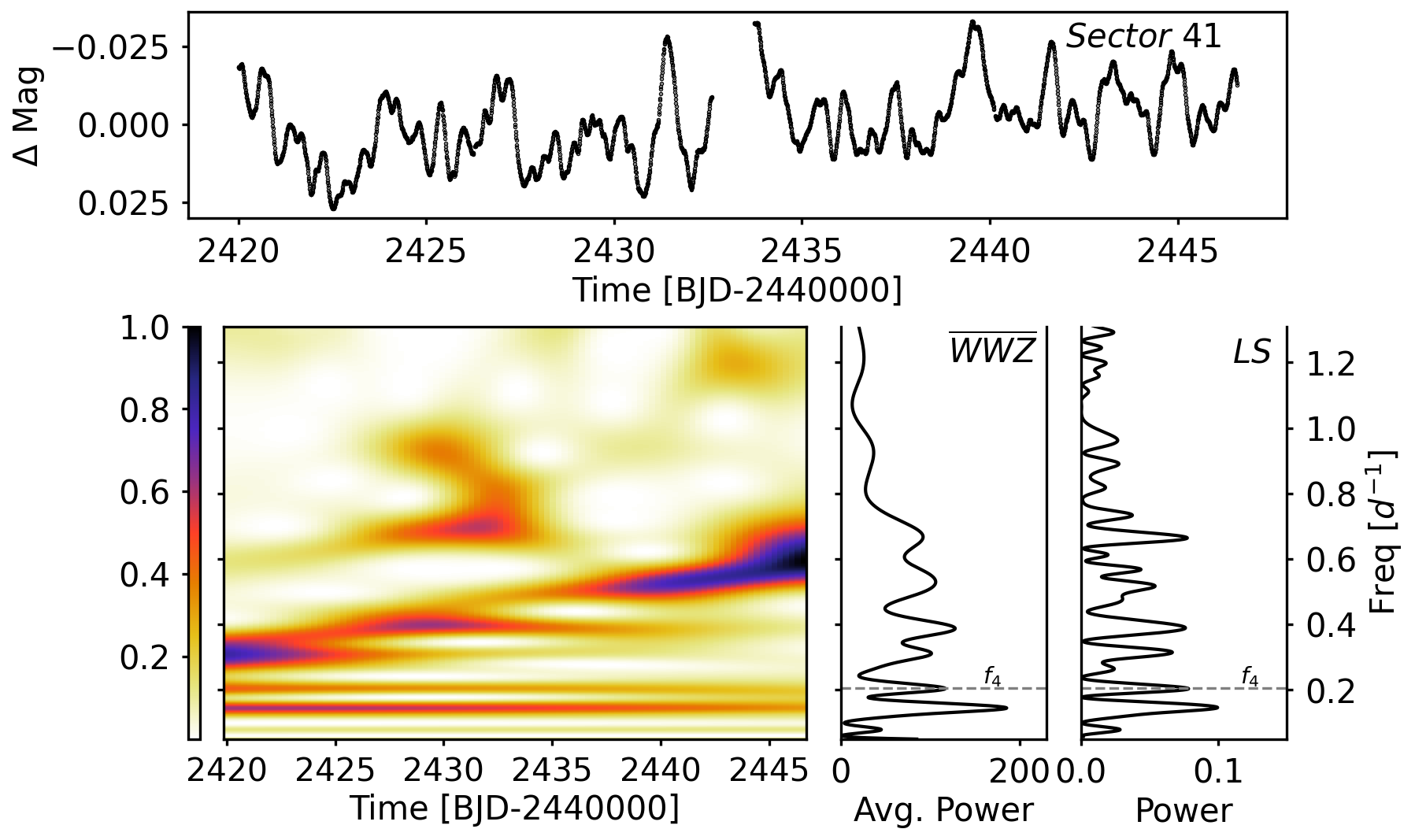}\\
    \vspace{10pt} \includegraphics[width=\columnwidth]{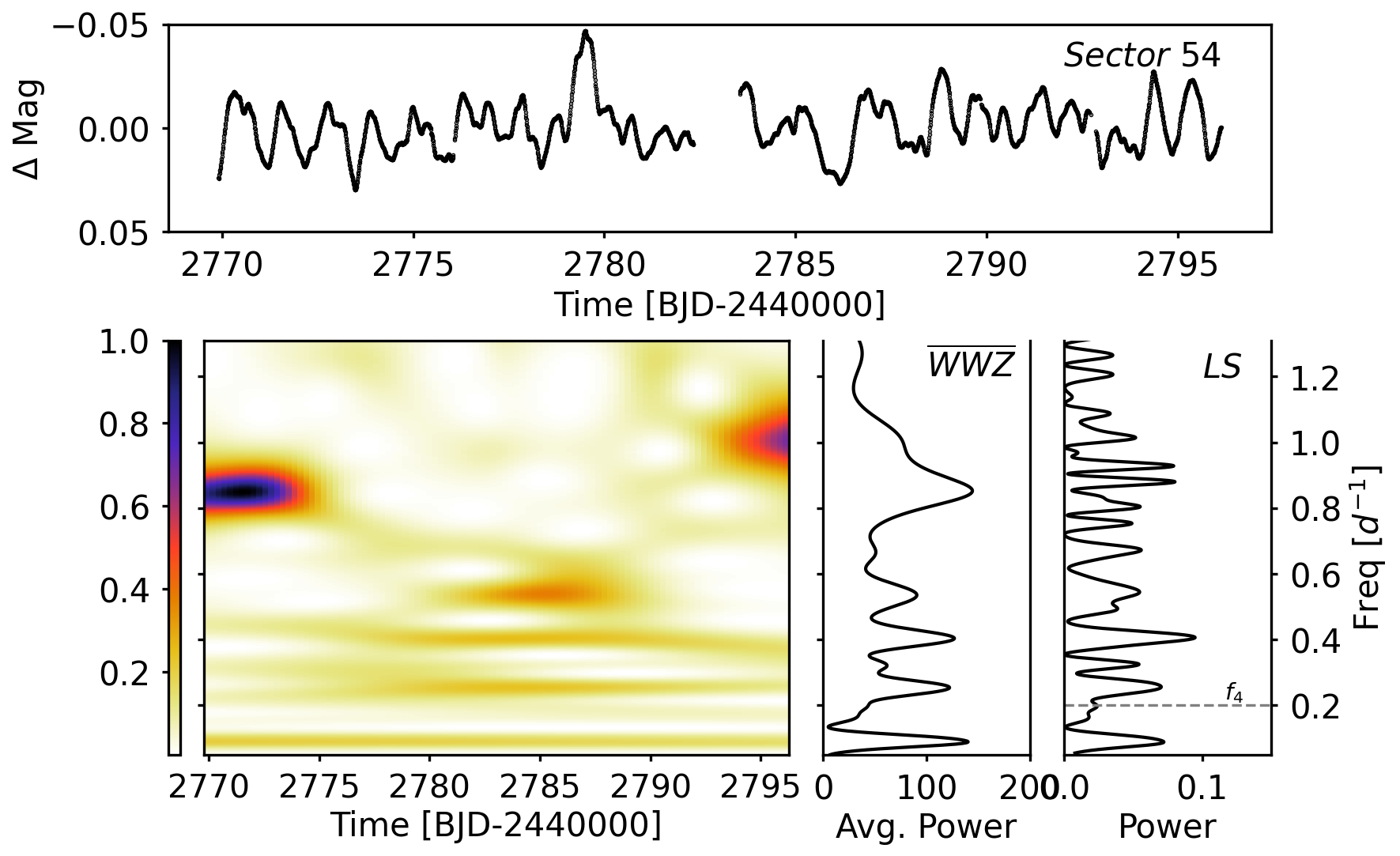}  
 \caption{\object{HD~188001}: TESS light curves from sectors $14$, $41$, and $54$, together with the associated WWZ scalograms, average power distributions, and LS periodograms. Independent frequencies are indicated with the dashed lines.} 
 \label{fig:lightcurve9sge}
\end{figure}


\begin{figure}[h]
 \centering
\includegraphics[width=\columnwidth]{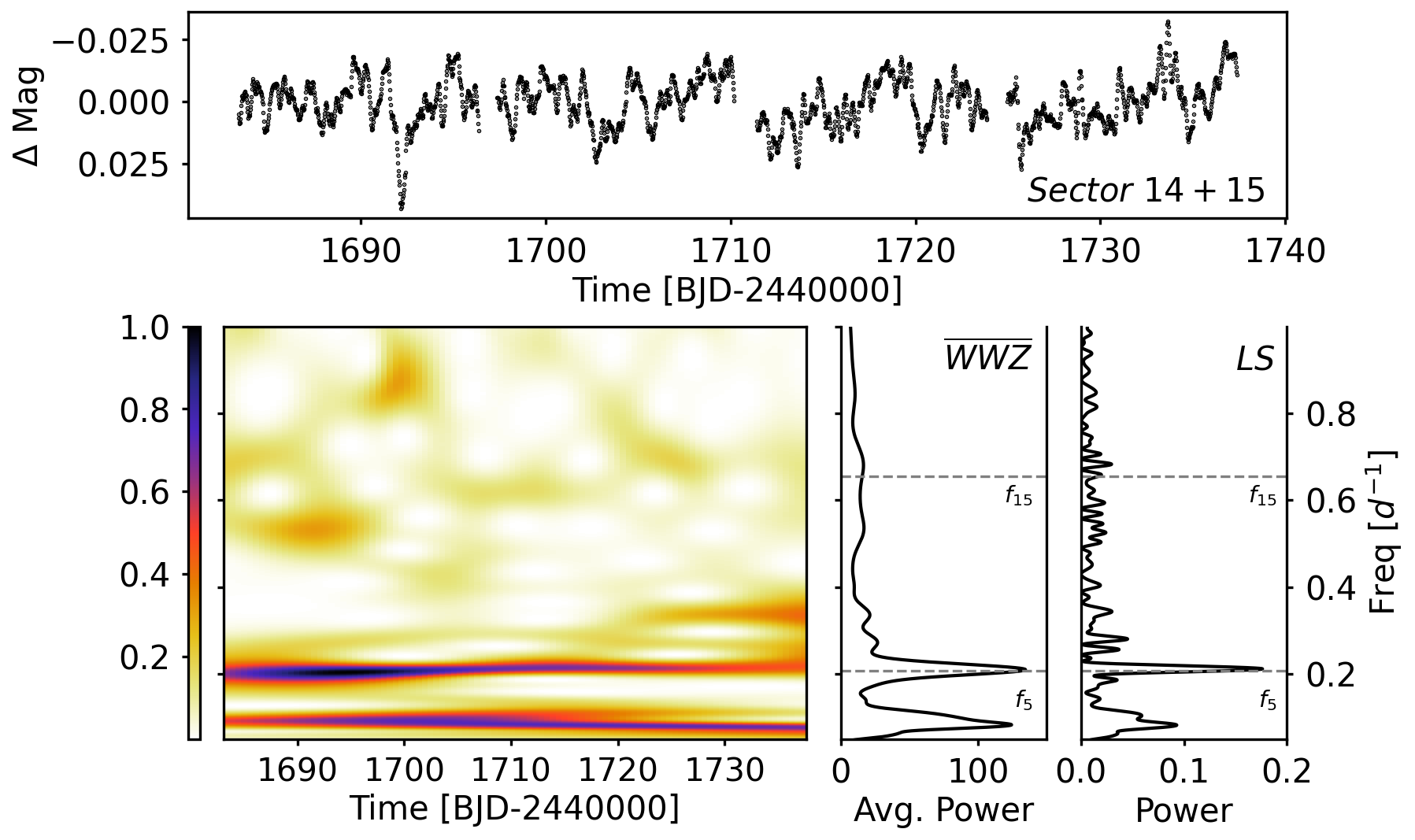}\\
\vspace{10pt}
\includegraphics[width=\columnwidth] {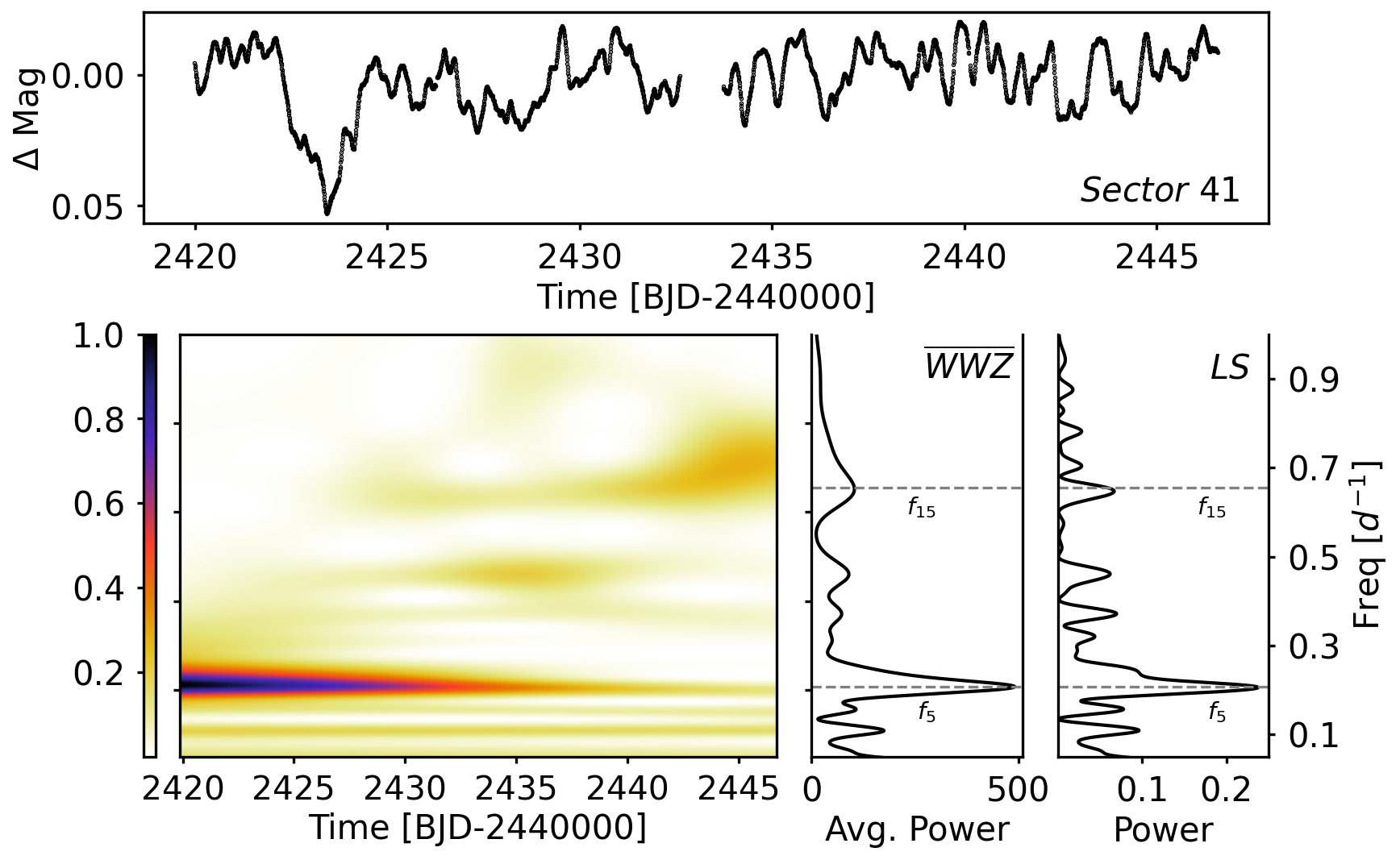}\\
\vspace{10pt}
\includegraphics[width=\columnwidth]{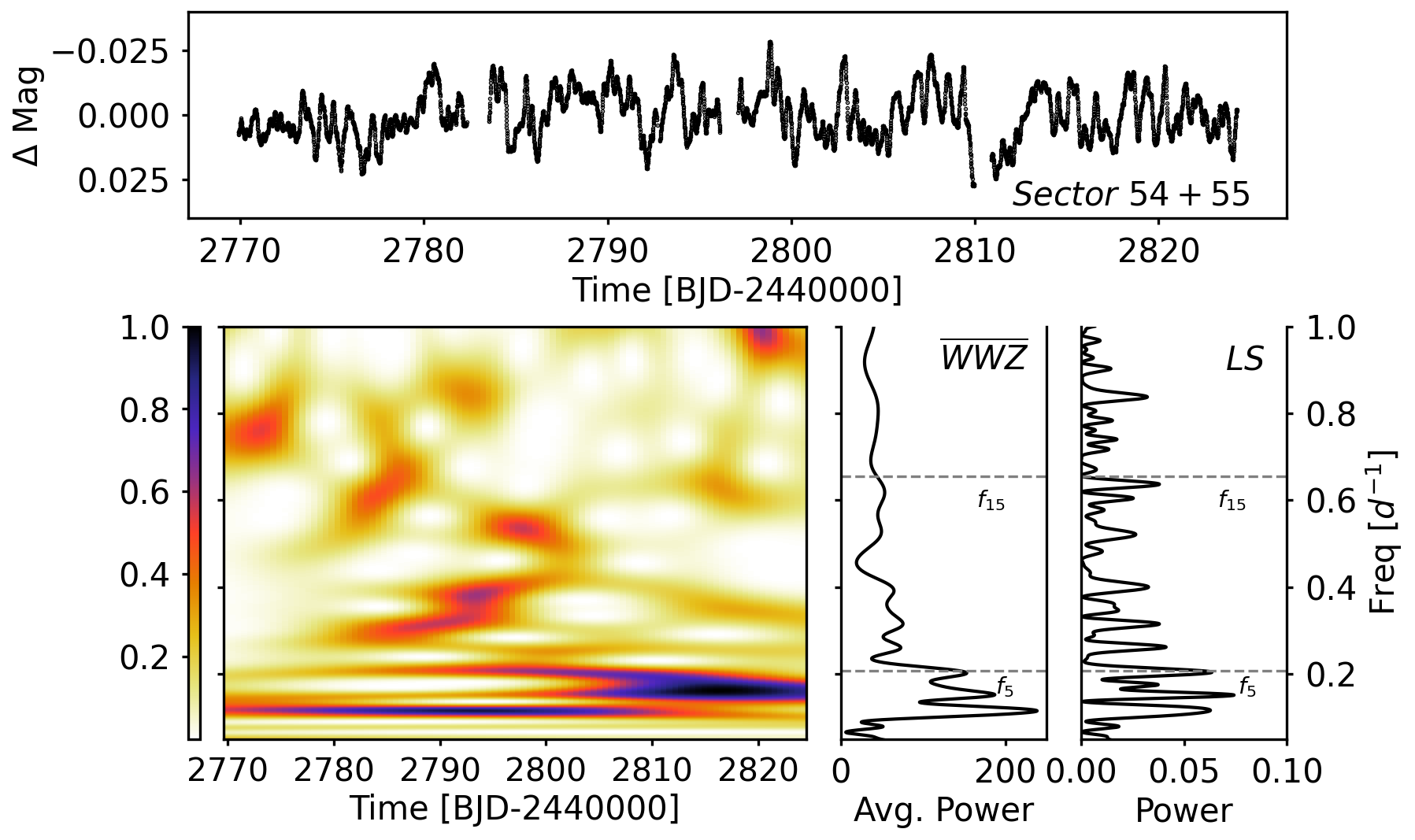}
\caption{\object{HD~192639}: TESS light curves for sectors $14$ and $15$, $41$, and $54$ and $55$, together with their corresponding WWZ scalograms, average power distributions, and LS periodograms. Independent frequencies are represented by the dashed lines.}
\label{fig_lightcurveHD192639}
\end{figure}


\begin{figure}[h!]
 \centering
\includegraphics[width=\columnwidth]{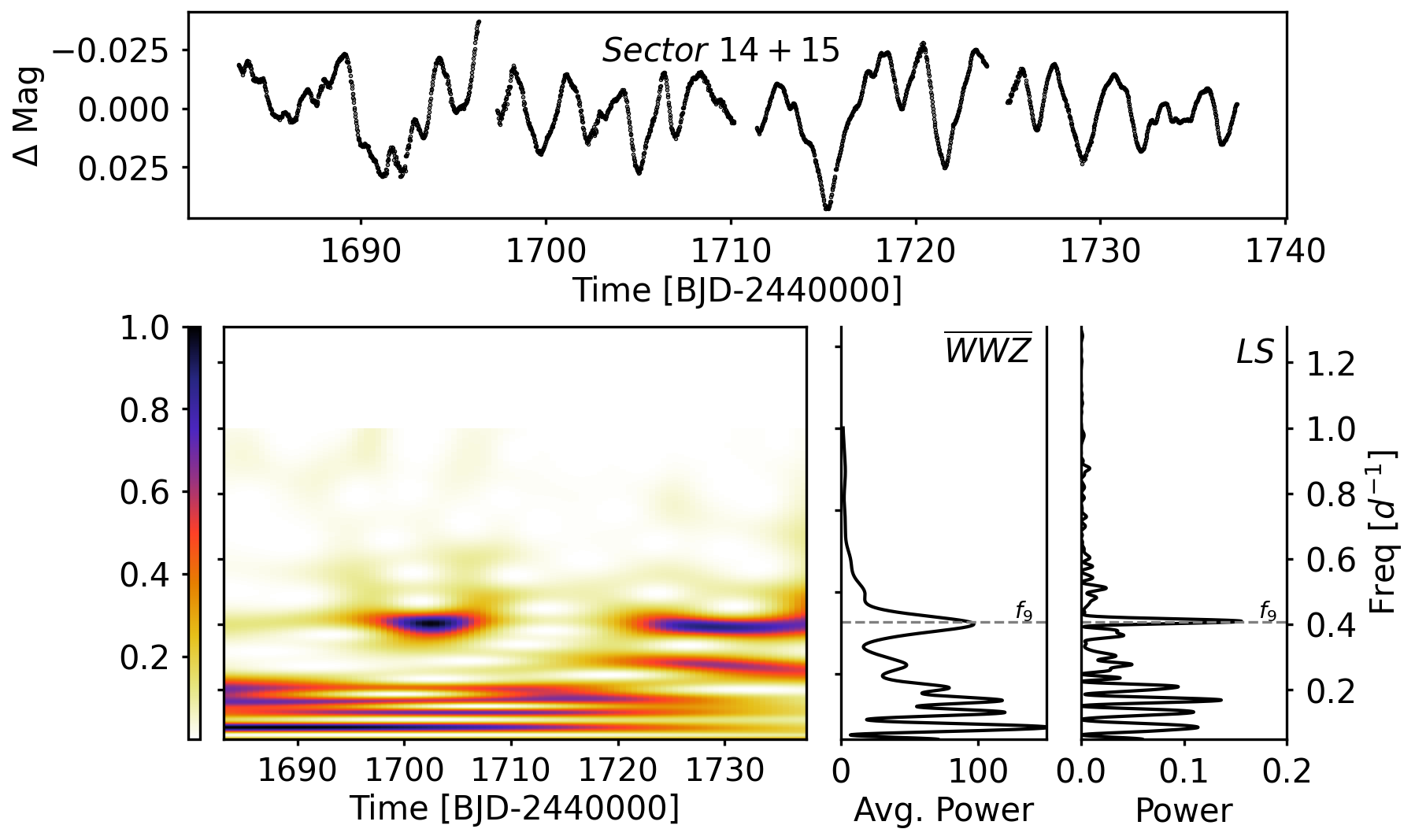}
   \vspace{10pt}
\includegraphics[width=\columnwidth] {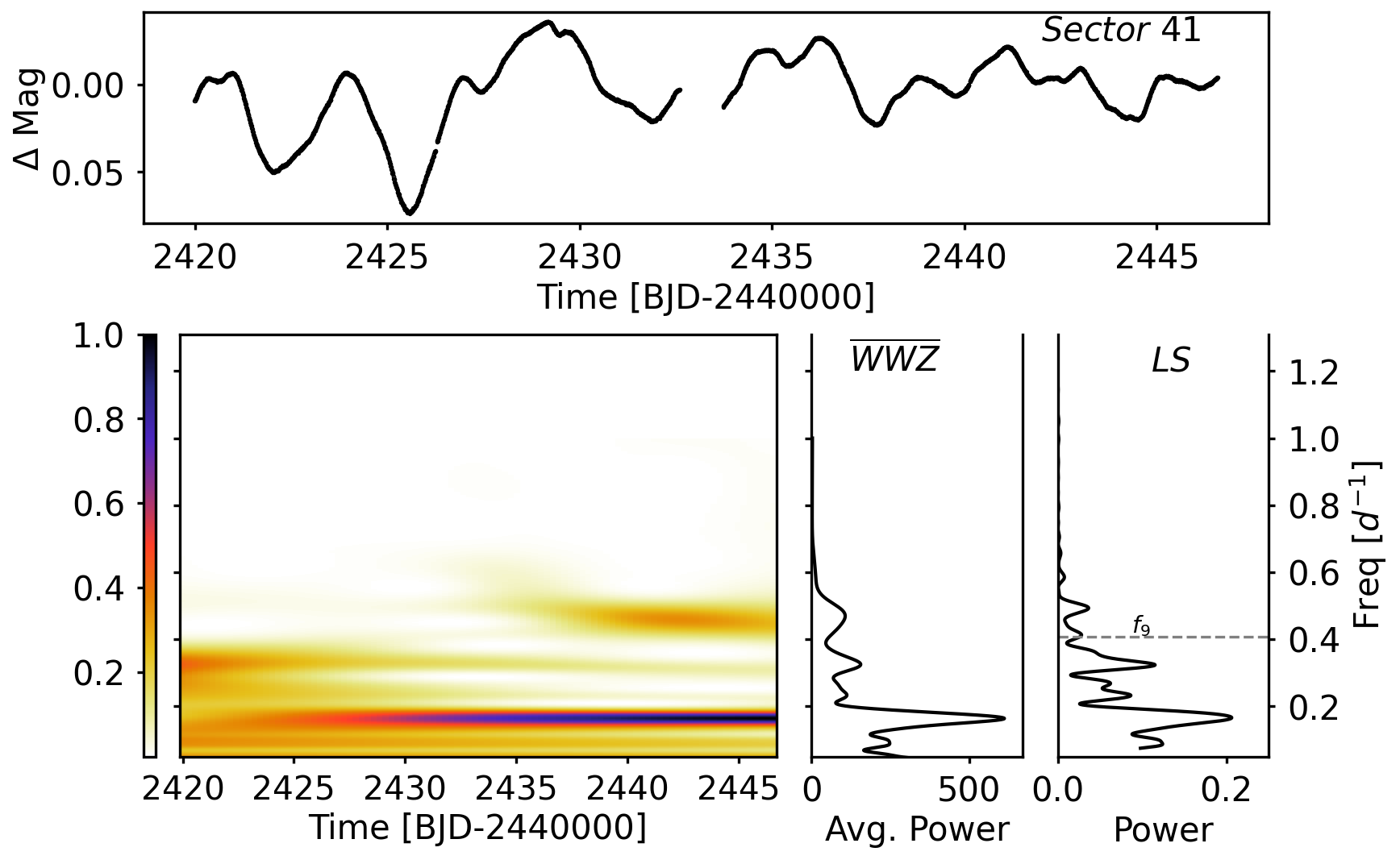}\\
   \vspace{10pt}     \includegraphics[width=\columnwidth]{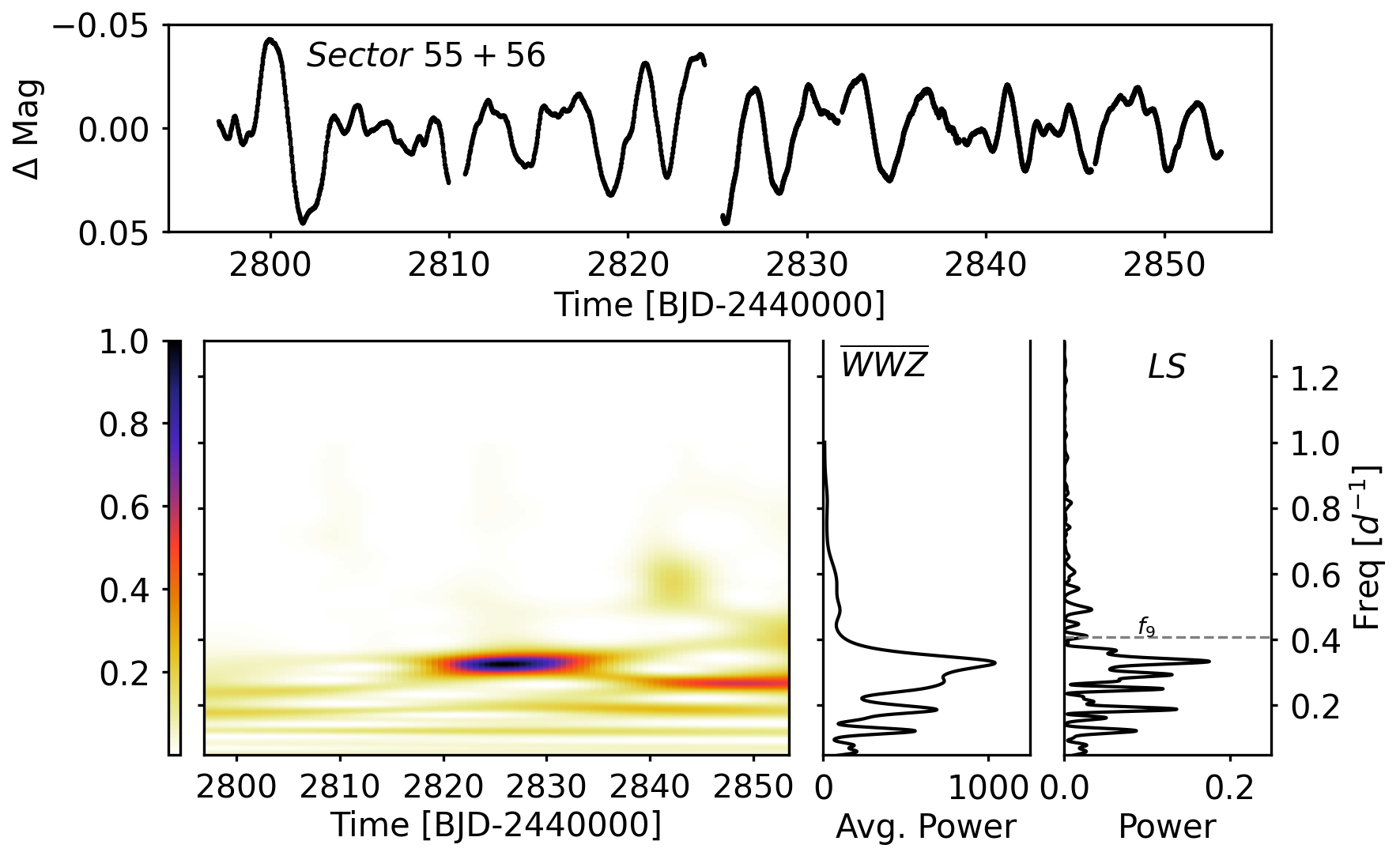} 
 \caption{\object{HD~195592}: TESS light curves for sectors $14$ and $15, 41$, and $55$ and $56$ are shown alongside their respective WWZ scalograms, average power distributions, and LS periodograms. Independent frequencies are indicated by a dashed line.}
 \label{fig:lightcurveHD195592}
\end{figure}

\onecolumn
 \section{Supplementary tables}
\label{Apex}

This appendix provides Tables~\ref{table:188001}, \ref{table:HD192639}, and \ref{table:HD195592} with details about the significant frequencies found for the objects \object{HD~188001}, \object{HD~192639}, and \object{HD~195592}, respectively.


\begin{table*}[h]
\caption{Significant frequencies for \object{HD~188001} in sectors 14, 41, and 54.}
\label{table:188001}
\centering
\newcolumntype{Y}{>{\centering\arraybackslash}X}
\newcolumntype{P}{>{\centering\arraybackslash}p{2.8cm}}
\begin{tabularx}{0.9\textwidth}{Y | Y Y | Y Y | Y Y | P | Y}
\hline
\hline
$Freq$ & \multicolumn{2}{c}{Sector 14} & \multicolumn{2}{c}{Sector 41} & \multicolumn{2}{c}{Sector 54} & $Average~f \pm \sigma$ & $Period$ \\
 & $LS$ & $WWZ$ & $LS$ & $WWZ$ & $LS$ & $WWZ$ & [d$^{-1}$] & [d]  \\
\hline
$f_{1}$  & $0.0746[2]$  & $0.0750[2]~$ & $0.0808[7]$ & $0.0793[7]~$ & $0.0907[3]$ & $0.0890[2]$~ & $0.0816 \pm 0.0063$ & $12.25$ \\
$f_{2}$  & $0.1118[3]$ & $0.1156[3]$~ & $\cdots$ & $\cdots$ & $\cdots$ &  $\cdots$ & $0.1137 \pm 0.0019$ & $8.79$\\ 
$f_{3}$  & $\cdots$ & $\cdots$ & $0.1448[1]$ & $0.1449[1]$ & $\cdots$ & $\cdots$ & $0.1449 \pm 0.0001$ & $6.90$\\ 
$f_{4}$  & $0.2004[1]$ & $0.1996[1]$ & $0.2051[5]$ & $0.2043[3]$ & $0.1977[7]$ & $\cdots$ & $0.2014 \pm 0.0028$ & $4.96$\\
$f_{5}$ & $0.2692[7]$ & $0.2596[6]$ & $\cdots$ & $\cdots$ & $0.2557[4]$ & $0.2543[4]$ & $0.2597 \pm 0.0058$ & $3.85$ \\
$f_{6}$ & $\cdots$ & $\cdots$ & $0.3119[3]$ & $0.3113[5]$ & $\cdots$ & $0.3213[6]$ & $0.3148 \pm 0.0046$ & $3.18$\\
$f_{7}$ & $0.3841[4]$ & $0.3816[4]$ & $0.3902[2]$ & $0.3873[2]$ & $0.4064[1]$ & $0.4033[3]$ & $0.3922 \pm 0.0094$ & $2.55$\\  
$f_{8}$ & $0.5104[9]$ & $0.5066[5]$ & $0.5204[3]$ & $0.5303[4]$ & $\cdots$ & $0.5353[5]$ & $0.5206 \pm 0.0110$ & $1.92$ \\ 
$f_{9}$ & $\cdots$ & $\cdots$ & $0.6647[2]$ & $0.6673[6]$ & $0.6634[5]$ & $0.6563[7]$ & $0.6629 \pm 0.0041$ & $1.51$ \\ 
$f_{10}$ & $0.7405[6]$ & $0.7496[8]$ & $\cdots$ & $\cdots$ & $\cdots$ & $\cdots$ & $0.7451 \pm 0.0046$ & $1.34$ \\
$f_{11}$ & $0.8715[5]$ & $0.8646[9]$ & $\cdots$ & $\cdots$ & $0.8573[6]$ & $0.8523[1]$ & $0.8614 \pm 0.0073$ & $1.16$ \\ 
$f_{12}$ & $\cdots$ & $\cdots$ & $0.9418[7]$ & $0.9243[8]$ & $0.9279[2]$ & $\cdots$ & $0.9313 \pm 0.0075$ & $1.07$ \\ 
$f_{13}$ & $1.0789[8]$ & $1.0796[7]$ & $\cdots$ & $\cdots$ & $\cdots$ & $\cdots$ & $1.0793 \pm 0.0003$ & $0.93$ \\ 
\bottomrule
\end{tabularx}
\tablefoot{Frequency values are in ascending order. The order of detection of the frequencies, according to their power, is indicated in brackets. The average frequency in each row, together with the standard deviation, is displayed in column~$5$. The period corresponding to the average frequency is in column~$6$.}
\end{table*}


\begin{table*}[h]
\caption{Significant frequencies for \object{HD~192639} for the combined sectors $14$ and $15$,  sector $41$, and sectors $54$ and $55$.}
\label{table:HD192639}
\centering
\newcolumntype{Y}{>{\centering\arraybackslash}X}
\newcolumntype{P}{>{\centering\arraybackslash}p{2.8cm}}
\newcolumntype{Q}{>{\centering\arraybackslash}p{2cm}}
\begin{tabularx}{0.9\textwidth}{Y| Y Y| Y Y| Y Y |P | Y}
\hline
\hline
$Freq$& \multicolumn{2}{c}{Sectors 14 \& 15} & \multicolumn{2}{c}{Sector 41} & \multicolumn{2}{c}{Sectors 54 \& 55} & $Average~w\pm\sigma$ & $Period$ \\
& \textit{LS} & \textit{WWZ} & \textit{LS} & \textit{WWZ} & \textit{LS} & \textit{WWZ} & [d$^{-1}$] & [d] \\
\hline
$f_1$ & $0.0810[2]$ & $0.0828[2]$ & $\cdots$ & $\cdots$ & $0.0842[18]$ & $\cdots$ & $0.0827\pm0.0016$ & $12.10$ \\
$f_2$ & $\cdots$ & $\cdots$ & $0.1083[6]$ & $0.1090[3]$ & $0.1105[2]$ & $0.1150[1]$ & $0.1107\pm0.0030$ & $9.03$ \\
$f_3$ & $0.1415[3]$ & $\cdots$ & $\cdots$ & $0.1570[4]$ & $0.1513[1]$ & $0.1522[2]$ & $0.1505\pm0.0065$ & $6.65$  \\
$f_4$ & $0.1889[8]$ & $\cdots$ & $0.1726[6]$ & $\cdots$ & $0.1852[15]$ & $\cdots$ & $0.1822\pm0.0085$ & $5.49$ \\
$f_5$ & $0.2116[1]$ & $0.2099[1]$ & $0.2056[1]$ & $0.2080[1]$ & $0.2056[3]$ & $0.2020[3]$ & $0.2071\pm0.0035$ & $4.83$ \\
$f_6$ & $0.2535[7]$ & $\cdot$ & $\cdots$ & $\cdots$ & $0.2636[6]$ & $0.2605[5]$ & $0.2592\pm0.0052$ & $3.86$\\
$f_7$ & $0.2890[11]$ & $0.2719[3]$ & $\cdots$ & $\cdots$ & $\cdots$ & $\cdots$ & $0.2805\pm0.0121$ & $3.57$\\
$f_8$ &$\cdots$ & $\cdots$ & $\cdots$ & $\cdots$ & $0.3167[4]$ & $0.3156[4]$ & $ 0.3162\pm0.0008$ & $3.16$ \\
$f_{9}$ & $0.3435[4]$ & $\cdots$ & $\cdots$ & $\cdots$ &  $\cdots$ & $\cdots$ & $0.3435\pm0.0000$ & $2.91$ \\
$f_{10}$ & $0.3783[15]$ & $\cdots$ & $0.3714[3]$ & $\cdots$ & $0.3980[5]$ & $0.3911[6]$ & $0.3847\pm0.0120$ & $2.60$\\
$f_{11}$ & $0.4760[12]$ & $\cdots$ & $0.4598[4]$ & $0.4600[6]$ & $0.4822[16]$ & $\cdots$ & $0.4695\pm0.0114$ & $2.13$\\
$f_{12}$ & $0.5242[6]$ & $\cdots$ & $0.5277[11]$ & $\cdots$ & $0.5179[9]$ & $\cdots$ & $0.5233\pm0.0050$ & $1.91$ \\
$f_{13}$ & $0.5697[18]$ & $\cdots$ & $\cdots$ & $\cdots$ & $0.5797[11]$ & $\cdots$ & $0.5747\pm0.0071$ & $1.74$ \\
$f_{14}$ & $0.5961[10]$ & $\cdots$ & $\cdots$ & $\cdots$ & $0.6001[12]$ & $\cdots$ & $0.5981\pm0.0029$ & $1.67$ \\
$f_{15}$ & $0.6826[5]$ & $\cdots$ & $0.6447[2]$ & $0.6540[5]$ & $0.6362[8]$ & $\cdots$ & $0.6544\pm0.0202$ & $1.53$ \\
$f_{16}$ & $0.7396[14]$ & $\cdots$ & $0.7113[8]$ & $\cdots$ & $0.7403[10]$ & $\cdots$ & $0.7304\pm0.0165$ & $1.37$ \\
$f_{17}$ & $0.7717[17]$ & $\cdots$ & $0.7826[9]$ & $\cdots$ & $0.7800[14]$ & $\cdots$ & $0.7781\pm0.0057$ & $1.29$ \\
$f_{18}$ & $0.8470[9]$ & $\cdots$& $\cdots$ & $\cdots$ & $0.8380[7]$ & $\cdots$ & $0.8425\pm0.0063$ & $1.19$ \\
$f_{19}$ & $0.8979[16]$ & $\cdots$ & $0.8801[7]$ & $\cdots$ & $0.8640[17]$ & $\cdots$ & $0.8807\pm0.0170$ & $1.13$\\ 
$f_{20}$ & $0.9354[13]$ & $\cdots$ & $0.9360[10]$ & $\cdots$ & $0.9019[13]$ & $\cdots$ & $0.9244\pm0.0195$ & $1.08$ \\
\hline
\end{tabularx}
\tablefoot{Details on the columns are given in Table~\ref{table:188001}.}
\end{table*}


\begin{table*}[h!]
\caption{Significant frequencies for \object{HD~195592} for each sector, the combined sectors $14$ and $15$, single sector $41$ and sectors $55$ and $56$.}
\label{table:HD195592}
\centering
\newcolumntype{Y}{>{\centering\arraybackslash}X}
\newcolumntype{P}{>{\centering\arraybackslash}p{2.3cm}}
\newcolumntype{Q}{>{\centering\arraybackslash}p{1cm}}
\begin{tabularx}{0.9\textwidth}{Y|YY|YY| Y Y |P |Y}
\hline
\hline
$Freq$& \multicolumn{2}{c}{Sectors 14 \& 15} & \multicolumn{2}{c}{Sector 41} & \multicolumn{2}{c}{Sectors 55 \& 56} & $Average~f\pm\sigma$ & $Period$ \\

& \textit{LS} & \textit{WWZ} & \textit{LS} & \textit{WWZ} & \textit{LS} & \textit{WWZ} & [d$^{-1}$] & [d] \\
\hline
$f_1$    & $0.0474[6]$  & $0.0487[6]$ & $\cdots$     & $\cdots$    & $0.0530[16]$ & $0.0596[5]$ & $0.0522\pm0.0055$ & $19.16$ \\
$f_2$    & $0.0858[3]$  & $0.0859[3]$ & $0.0819[3]$  & $0.0896[9]$ & $0.0868[10]$ & $0.0782[6]$ & $0.0847\pm0.0040$ & $11.81$ \\
$f_3$    & $0.1291[4]$  & $0.1316[2]$ & $0.1133[2]$  & $\cdots$    & $0.1206[6]$ & $0.1221[4]$ & $0.1233\pm0.0073$ & $8.11$ \\
$f_4$    & $\cdots$     & $\cdots$    & $\cdots$     & $\cdots$    & $0.1534[7]$ & $\cdots$ & $0.1534\pm0.0000$ & $6.52$ \\
$f_5$    & $0.1687[2]$  & $0.1681[3]$ & $0.1686[1]$  & $0.1640[1]$ & $0.1732[9]$ & $\cdots$ & $0.1685\pm0.0033$ & $5.93$ \\
$f_6$    & $\cdots$     & $\cdots$    & $\cdots$     & $\cdots$    & $0.1879[2]$ & $0.1868[3]$ & $0.1873\pm0.0008$ & $5.34$ \\
$f_7$    & $0.2083[5]$  & $0.2065[5]$ & $0.2136[12]$ & $\cdots$    & $0.2162[17]$ & $\cdots$ & $0.2111\pm0.0045$ & $4.74$\\
$f_8$    & $\cdots$     & $\cdots$    & $0.2571[5]$  & $0.2337[4]$ & $0.2489[4]$ & $\cdots$ & $0.2465\pm0.0119$ & $4.06$\\
$f_9$    & $0.2662[9]$  & $0.2760[7]$ & $\cdots$     & $\cdots$    & $0.2632[21]$ & $0.2741[2]$ & $0.2699\pm0.0061$ & $3.71$ \\
$f_{10}$ & $0.2803[8]$  & $\cdots$    & $\cdots$     & $\cdots$    & $\cdots$ & $\cdots$ & $0.2803\pm0.0000$ & $3.57$\\
$f_{11}$ & $0.2923[13]$ & $\cdots$    & $0.2942[8]$  & $\cdots$    & $0.2937[3]$ & $\cdots$ & $0.2934\pm0.0010$ & $3.41$ \\
$f_{12}$ & $0.3129[10]$ & $\cdots$    & $\cdots$     & $\cdots$    & $0.3169[11]$ & $\cdots$ & $0.3149\pm0.0028$ & $3.18$ \\
$f_{13}$ & $\cdots$     & $\cdots$    & $0.3299[4]$  & $0.3252[3]$ & $0.3336[1]$ & $0.3290[1]$ & $0.3294\pm0.0034$ & $3.04$ \\
$f_{14}$ & $\cdots$     & $\cdots$    & $\cdots$     & $\cdots$    & $0.3563[5]$ & $\cdots$ & $0.3563\pm0.0000$ & $2.81$\\
$f_{15}$ & $0.3676[7]$  & $\cdots$    & $0.3626[6]$  & $\cdots$    & $0.3698[22]$ & $\cdots$ & $0.3667\pm0.0037$ & $2.73$ \\
$f_{16}$ & $\cdots$     & $\cdots$    & $\cdots$     & $\cdots$    & $0.3886[13]$ & $\cdots$ & $0.3886\pm0.0000$ & $2.57$ \\
$f_{17}$ & $0.4097[1]$  & $0.4010[4]$ & $0.4153[9]$  & $\cdots$    & $0.4078[20]$ & $\cdots$ & $0.4084\pm0.0059$ & $2.45$\\
$f_{18}$ & $\cdots$     & $\cdots$    & $0.4552[10]$ & $0.4694[5]$ & $\cdots$ & $\cdots$ & $0.4623\pm0.0101$ & $2.16$\\
$f_{19}$ & $0.4820[12]$ & $0.4907[8]$ & $0.4900[7]$  & $\cdots$    & $0.4912[8]$ & $0.4891[7]$ & $0.4886\pm0.0038$ & $2.05$ \\
$f_{20}$ & $0.5129[11]$ & $\cdots$    & $0.5301[11]$ & $\cdots$    & $\cdots$ & $\cdots$ & $0.5215\pm0.0121$ & $1.92$ \\
$f_{21}$ & $0.5488[14]$ & $\cdots$    & $\cdots$     & $\cdots$    & $0.5529[12]$ & $0.5646[8]$ & $0.5554\pm0.0082$ & $1.80$ \\
$f_{22}$ & $\cdots$     & $\cdots$    & $\cdots$     & $\cdots$ & $0.5808[18]$ & $\cdots$ & $0.5808\pm0.0000$ & $1.72$\\
$f_{23}$ & $\cdots$     & $\cdots$    & $\cdots$     & $\cdots$ & $0.6076[14]$ & $\cdots$ & $0.6076\pm0.0000$ & $1.65$ \\
$f_{24}$ & $\cdots$     & $\cdots$    & $\cdots$     & $\cdots$ & $0.6474[15]$ & $\cdots$ & $0.6474\pm0.0000$ & $1.54$\\
$f_{25}$ & $\cdots$     & $\cdots$    & $\cdots$     & $\cdots$ & $0.7166[19]$ & $\cdots$ & $0.7166\pm0.0000$ & $1.40$ \\
$f_{26}$ & $\cdots$     & $\cdots$    & $\cdots$     & $\cdots$ & $0.8439[23]$ & $0.8163[9]$ & $0.8301\pm0.0195$ & $1.20$\\
\hline
\end{tabularx}
\tablefoot{Details on the columns are given in Table~\ref{table:188001}.}
\end{table*}
\clearpage

\section{Analysis of selected average frequencies}

\begin{table*}[h]
\centering
\caption{\object{HD 192639}: Selected frequencies from Table~\ref{table:HD192639}.}
\label{HD192639_analysis}
\begin{tabular}{rlrcrr|cc|rcc}
\hline
\hline
$ID$&\multicolumn{5}{c}{\it Observed values}~& \multicolumn{2}{c}{$Theoretical$} &\multicolumn{3}{c}{$Comments$}\\
& \hspace{1cm}$f_{i}$ & ~~~~$f_{i}-f_{i-1}$ && $P$ & $\Delta P$ &$f_{e}$ & $f-f_{e}$&& \\
 &\hspace{0.8cm}[d$^{-1}$]&[d$^{-1}$]&&[d]& [d]& ~[d$^{-1}$]&~[d$^{-1}$]&&\\
\hline
$1$ &$0.0827\pm0.0016$ &&& $12.09$ & $3.06$ &$0.0829$ & $-0.0002$ & $\frac{3}{2}~\Delta~f$ && $(2,2)$\\
$2$&$\mathbf{0.1107\pm0.0030}$&$0.0280$&&$9.03$ &$2.39$&$0.1037$&$0.0070$&$2\,\Delta f$&$f_5-2\,\Delta f$&$(4,2)$\\
$3$& $\mathbf{0.1505\pm0.0065}$&  $0.0398$    & &$6.64$ &$1.16$ &$0.1556$&$-0.0051$ &$3\,\Delta f$& $f_5-\Delta f$& $g-mode$ - $(3,3)$\\
$4$& $0.1822\pm0.0085$ & $0.0317$&&$5.49$ &$0.66$ & $0.1815$&$0.0007$&$\frac{7}{2}\,\Delta f$&&  $g-mode$ - $(7,3)$\\
$5$& $\mathbf{0.2071\pm0.0035}$    &$0.0249$& &$4.83$ &$0.97$ &$0.2074$& $-0.0003$& 
 $4\,\Delta f$ & $f_5$ & $g-mode$\\
$6$&$\mathbf{0.2592\pm0.0052}$& $0.0521$&&$3.86$& $0.29$& $0.2593$ &$-0.0001$&$5\,\Delta f$&$f_5+\Delta f$& $g-mode$\\
$7$&$0.2805\pm0$&$0.0121$&&$3.57$&$0.40$ &$0.2852$&$-0.0047$&$\frac{11}{2}\,\Delta f$&& $g-mode$\\
$8$ & $\mathbf{0.3162\pm0.0008}$& $0.0356$ && $3.16$ &$0.25$&  $0.3111$ & $0.0050$ & $6\,\Delta f$ & $f_5+ 2\,\Delta f$ &$g-mode$\\
$9$&$0.3435\pm0$&$0.0274$&&$2.91$&$0.31$&$0.3370$&$0.0060$&$\frac{13}{2}\Delta f$& & $g-mode$\\
$10$ & $0.3847\pm0.0008$ & $0.0412$ && $2.60$ &$0.47$ & $0.3889$&$-0.0042$& $\frac{17}{2}\,\Delta f$ &&$g-mode$\\
$11$&$\mathbf{0.4695\pm0.0045}$&$0.0848$&&$2.13$&$0.22$&$0.4667$&$0.0028$ &$9\,\Delta f$&&$g-mode$\\
$12$&$\mathbf{0.5233\pm0.0119}$& $0.0538$ && $1.91$&$0.17$& $0.5185$& $-0.0048$ & $10\,\Delta f$&&$g-mode$\\
$13$&$\mathbf{0.5747\pm0.0061}$&$0.0514$&&$1.74$&$0.07$&$0.5704$&$0.0040$&$11\,\Delta f$&&\\
$14$ &$0.5981\pm0$ & $0.0234$& &$1.67$ &$0.14$ & $0.5963$ & $-0.0018$& $\frac{23}{2}\,\Delta f$&&$g-mode$\\
$15$ &$\mathbf{0.6544\pm0.001}$ & $0.0563$&&$1.53$ &$0.16$ &$0.6481$&$-0.0063$ & $\frac{25}{2}\,\Delta f$&&\\
$16$& $0.7308\pm0.0028$& $0.0760$ &&$1.37$ &$0.08$&$0.7259$&$-0.0045$&$14\,\Delta f$&&\\
$17$&$0.7781\pm0.0034$&$0.0477$&&$1.29$&$0.10$&$0.7778$ &$-0.0003$ & $15\,\Delta f$ &&$g-mode$\\
\hline
\end{tabular}
\tablefoot{Column 2 lists the average observed frequency ($f$). The difference between consecutive rows is given in column 3. The obtained period is in column 4. Using a mean-square algorithm, we estimate an average spacing of $\Delta f = 0.05185$~d$^{-1}$ between the highlighted frequencies (given in bold font). The identified $g$ modes appear to cluster around integer and half-integer multiplets of $\Delta f$ ($f_e= \frac{m}{2} \Delta f$ with $m=1$ to $30$; see columns 6, 7, and 8). The frequency $f_5$ is taken as a reference for a rotational splitting. Tentative $r$ modes, labelled by $(\ell,m)$, are indicated in the last column.}
\end{table*}

\begin{table*}
\caption{\object{HD 195592}: Selected frequencies from Table~\ref{table:HD195592}.}
\label{analysis_HD195592}
\begin{tabular}{rlccrr|cc|rcl}
\hline
\hline
$ID$ &\multicolumn{5}{c} {$Observed\, values$} & \multicolumn{2}{c}{$Theoretical$} & \multicolumn{3}{c}{$Comments$}\\
  & \hspace{1cm} $f_i$ & $f_{i}-f_{i-1}$ & $f_{i}-f_{i-2}$ & $P$ &$\Delta~P$& $f_e$ & $fi-f_e$ &&\\
  &\hspace{0.8cm}[d$^{-1}$]&[d$^{-1}$]&[d$^{-1}$]&[d]& [d]& ~[d$^{-1}$]&~[d$^{-1}$]&&\\
\hline
 $1$ & $\mathbf{0.0847\pm0.0029}$ &         &                  & $11.81$ &$3.70$& $0.0818$ & $0.0028$ &$\Delta f$&& \\
 $2$ & $0.1233\pm0.0073$          &$0.0399$ & $\mathbf{0.0825}$& $8.11$ &$2.18$  &&&$\frac{3}{2}\,\Delta f$&&\\
 $3$ & $\mathbf{0.1685\pm0.0033}$ & $0.0426$ &                  & $5.93$ &$1.19$ &$0.1637$&$0.0048$&$2\,\Delta f$&& $g-mode$ \\
 $4$ & $0.2111\pm0.0045$          &$0.0426$ &$\mathbf{0.0780}$ &$4.74$ &$0.68$& 
 &       &      $\frac{5}{2}\,\Delta f$ &     & $g-mode$ - $(2,2)$\\
 $5$ & $\mathbf{0.2465\pm0.0119}$ &$0.0354$ &                  &$4.06$ &$0.49$&$0.2455$& $0.0009$& $3\,\Delta f$& &  $g-mode$\\ 
 $6$& $0.2803\pm0$                &$0.0338$ &$\mathbf{0.0829}$ &$3.57$ &$0.53$&&&$\frac{7}{2}\,\Delta f$&& $g-mode$ - $(3,2)$\\
 $7$& $\mathbf{0.3294\pm0.0034}$  &$0.0491$ &                  &$3.04$ &$0.31$ & $0.3274$&$0.0020$&$4\,\Delta f$ &$f_9-\Delta f$ & $g-mode$\\
 $8$& $0.3667\pm0.0037$           &$0.0373$ &$\mathbf{0.0790}$ &$2.73$ &$0.28$& &&$\frac{9}{2}\,\Delta f$&&$g-mode$\\
 $9$& $\mathbf{0.4084\pm0.0059}$  &$0.0417$ &               &$2.45$ &$0.29$ &$0.4092$&$-0.0008~~$&$5\,\Delta f$&$f_9$&$g-mode$ - $(3,3)$\\
$10$& $0.4626 \pm 0.0101$                &0.0542 &$\mathbf{0.0802}$ &$2.16$ &$0.11$ &&&$\frac{11}{2}\,\Delta f$&& $g-mode$ - $(5,3)$\\
$11$ & $\mathbf{0.4886\pm0.0038}$  & $0.026$ &                  & $2.05$& $0.13$& $0.4911$&$-0.0025~~$&$6\,\Delta f$&$f_9+\Delta f$&\\
$12$& $0.5215\pm0.0121$ &             $0.0329$ &$\mathbf{0.0922}$ &$1.92$ &$0.2$&&&$\frac{13}{2}\,\Delta f$&&$g-mode$\\
$13$& $\mathbf{0.5808\pm0}$ & $0.0593$ &&$1.72$ &$0.07$&$0.5729$&$0.0078$&$7\,\Delta f$&&$g-mode$\\ 
$14$ &$0.6076\pm0$& $0.0268$ &$\mathbf{0.0666}$&$1.65$&$0.11$&&&$\frac{15}{2}\,\Delta f$&& $g-mode$ - $(5,4)$\\
$15$& $\mathbf{0.6474\pm0}$ &$0.0398$ && $1.54$&$0.14$&$0.6548$&$0.0074$&$8\,\Delta f$&&$g-mode$ \\
$16$ &$\mathbf{0.7166\pm0}$ && $\mathbf{0.0692}$ &$1.40$ &$0.2$& $0.7366$ &$-0.0200~~$& $9\,\Delta f$ &&\\
$17$ & $\mathbf{0.8301\pm0.0195}$ && $\mathbf{0.1135}$&$1.20$& &$0.8185$&$0.0116$& $10\,\Delta f$ && $g-mode$\\
\hline
\end{tabular}
\tablefoot{Column 2 lists the relevant average observed frequency ($f$). The frequency differences between rows separated by one and two positions are given in Columns 3 and 4, respectively. The obtained period is in column 4. The highlighted frequencies (in boldface) show an average spacing of $\Delta f = 0.08185$~d$^{-1}$.  The observed frequencies appear to cluster around integer and half-integer multiplets of $\Delta f$ ($f_e= \frac{m}{2} \Delta f$ with $m=1$ to $30$; see columns 6, 7, and 8). 
A clear rotational splitting pattern is observed among the $g$ modes, with frequency $f_9$ adopted as a representative case. The identified $g$ modes, along with candidate $r$ modes, are listed in the last column.}
\end{table*}

\end{appendix}
\end{document}